\documentclass[aps,prb,twocolumn,floatfix,superscriptaddress,showpacs,10pt]
{revtex4-1}
\usepackage{graphicx}
\usepackage{amssymb}
\usepackage{amsmath}
\makeindex

\newcommand\rs[1]{{\scriptscriptstyle\rm #1}}
\newcommand\llangle{{\langle\langle}}
\newcommand\rrangle{{\rangle\rangle}}

\begin{document}

\title{Lasing at half the Josephson frequency with exponentially long coherence
times}

\author{Frans Godschalk}

\author{Yuli V. Nazarov}
\affiliation{Kavli Institute of Nanoscience, Delft University of Technology,
P.O.  Box 5046, 2600 GA Delft, The Netherlands}

\date{November 2012}

\begin{abstract}
We describe a superconducting device capable of producing laser light in the
visible range at half the Josephson generation frequency, with the optical phase
of the light locked to the superconducting phase difference. An earlier
proposed device, the so called ``half-Josephson laser'' [Phys. Rev. Lett.
107,  073901 (2011)], cannot provide long coherence times, because of
spontaneous switchings between the emitter states. To circumvent this we
consider $N\gg1$ emitters driving an optical resonator mode. We derive a general
model that captures essential physics of such devices while not depending on
specific microscopic details. We find the conditions under which the coherence
times are exponentially long, thus surpassing the fundamental limitation on the
coherence times of common lasers. For this we study the noise in the device. In
particular, we are
interested in the rate of large fluctuations of the light field in the limit
where the typical fluctuations are small. The large fluctuations are
responsible for switching of the laser between stable states of radiation and
therefore determine the coherence time. 
\end{abstract}


\maketitle
All numerous degrees of freedom of a bulk superconductor are frozen at low
temperature. So its state is characterized by a single variable, the quantum
phase of the superconducting order parameter.\cite{tinkham} In principle, this
phase remains constant, that is coherent, for infinitely long time.
Another example of a highly coherent system is a laser. There a macroscopic
number of photons form a coherent state characterized by an optical phase and an
amplitude.\cite{scully} In contrast to superconductors the coherence time is
not infinite, but limited by the intrinsic noise in the device, caused by
spontaneous photon emission. This leads to a phase drift and thus to coherence
loss at typical time scales of\cite{scullylamb} $\tau_\text{dec}\simeq
n/\Gamma$,
with $n$ the number of
photons in the mode and $\Gamma$ the cavity escape rate. Coupling these two very
different systems may open up novel ways of controlling either of them, allowing
for manipulation of one with the other. 

Recently, a device has been proposed\cite{godschalk} that combines the coherence
of laser light with the coherence of the superconducting order parameter,
dubbed the ``half-Josephson laser'' (HJL). This device
was based on yet another recently proposed device, the ``Josephson light 
emitting diode''\cite{recher:10} (JoLED), which, it was shown, can be used for
quantum manipulation purposes.\cite{hassler} The light
in the
JoLED is generated by a biased semiconductor {\it p-n} junction realized on a
nanowire. To acquire a narrow emission spectrum of the junction, quantum dots
are defined at the {\it p-n} interface. Using nanowires an efficient coupling
can be achieved between semi- and superconductors.\cite{vanDam&Doh} By
connecting both ends of the described {\it p-n} semiconductor nanowire to
superconductors, superconductivity is induced at the quantum dot regions that
alter the quantum dot states. Electrons and holes jump from the leads to the 
quantum dots and recombine there, emitting a photon. Even though the JoLED is
biased with a voltage much larger than the superconducting gap, the
superconducting correlations are retained because any direct, nonradiative 
charge transfer is blocked by the potential barriers in the {\it
p-n} junction. The electron-hole recombination can happen in two ways. A
``blue'' photon at about the Josephson frequency, $\omega_J$, is emitted when
Cooper pairs from either side of the junction recombine. The phase of this
blue photon is locked to the superconducting phase difference. A process
resembling spontaneous parametric down-conversion\cite{spdc} allows for the
emission of two ``red'' photons in stead of a single blue one, at about half
the Josephson frequency, $\omega_J/2$. The JoLED is the basis for the
setup of the HJL. The latter exploits the
emission of red photons by the JoLED and amplifies it by stimulated
emission in an optical resonator. In contrast to the case of the JoLED, where
red photons are emitted spontaneously, the red photons in the resonant
mode of the HJL form a coherent state. The phase of this coherent state is
locked to the superconducting phase difference.

Lasing was found for the red emission with the optical phase locked to
the superconducting phase difference, where only two values of the optical phase
are allowed, having a phase difference of $\pi$. The importance of the phase
locking in this device is that it can result in exponentially long
laser coherence times. The idea to prolong coherence times by phase locking has
been thoroughly explored in the field of lasers.\cite{ohtsu} Using, for
instance, the technique of injection locking\cite{lang} one could pump a laser
with another one with longer coherence time, so that the pumped laser inherits
this longer coherence time. However, in that case the coherence time is still
limited by the
fundamental noise processes in the pump laser, with some
coherence time $\tau_\text{dec}$. The essential difference with the HJL is
that the laser light inherits the coherence time of the superconductors, which
is infinitely long. It will not become
infinitely long, though, because the fundamental noise in the laser is still
present and causes switchings to the lasing state with opposite phase, with
exponentially small probability. Hence, the coherence times can be exponentially
long.

Unfortunately, despite this promising prospect, the HJL does not exhibit these
exponentially long coherence times because switchings between quantum dot
states result in a shorter coherence time. The magnitude and phase of the laser
light depends on the actual quantum state of the JoLED in the HJL. Not every
state supports lasing however, and the ones that do give rise to various
magnitudes and phases of the laser light. Switchings between the quantum dot
states therefore cause decoherence of the HJL. The switching rates are
suppressed
because they occur in combination with the emission of off-resonant photons,
which is allowed because of the finite linewidth of the optical resonator. For
an optimal choice of parameters it is found that, despite this very different
decoherence mechanism, the coherence time is, $\tau_\text{dec}$, similar to
that of the common laser and much shorter than the envisioned exponentially long
coherence times.

The design for the HJL is based on a single quantum emitter, putting rather
high demands on the quality factor of the optical resonator. It was
proposed\cite{godschalk} to relax
these demands by considering a large number, $N\gg1$, emitters in a single
resonator, increasing proportionally the upper bound on the damping rate at
which lasing is achieved. As an added benefit, intensity fluctuations are
expected to reduce as $1/\sqrt{N}$, thus realizing a less fluctuating light
intensity. 

In this article, we thoroughly investigate the alternative idea, where, instead
of a single JoLED, we consider a large number of emitters forming a dipole
moment to drive the optical resonator mode.
The model that we introduce will be quite general as we will only consider
macroscopic quantities and leave out the microscopic details. 
The crucial aspect of the design, the coupling of the quantum states
to both superconducting leads, will of course 
remain, since it both drives the laser and induces the phase lock between
optical phase and superconducting phase difference. The high demands on the
quality factor of the resonator can be relaxed because it is driven by a big
number of emitters. 

We find that stationary lasing is possible with two equivalent values of the
optical phase, as in the HJL, and with exponentially long coherence times owing
to the phase lock. Rather surprisingly, the sole increase in number of emitters
does not directly guarantee the increased number of photons in the mode. The
average dipole moment of all emitters will turn out to be zero when the
emitters are coupled to the resonator mode only. Incoherent emission to other
modes is required to create a population
imbalance,\cite{populationinversion} resulting in a nonzero average dipole
moment. Owing to the
large number of emitters, single switchings will not cause decoherence as in the
HJL. Rather, they induce minor fluctuations of the dipole moment, transferring
to
detuning fluctuations in the resonator mode. These again transfer to
fluctuations of magnitude and phase of the lasing state. The detuning
fluctuations compete with the intrinsic fluctuations of the resonator mode. When
both are sufficiently small, the superconducting phase difference will remain
locked to a single value of the optical phase. The major decoherence mechanism
of the laser is then provided by large fluctuations, represented by the tails of
the noise distributions. These bring the optical field to the state with
opposite phase. Large switchings occur rarely, on exponentially long time
scales.
The importance of such a feature is hard to underestimate. Indeed, examining
these
processes we have found coherence times (surpassing the fundamental limit for
common lasers), scaling as $\tau_\text{dec}\sim \exp(n)$, up to some critical
value $n_s\gg1$, where it saturates. While the linewidth of the laser is
directly related to this coherence time, there are large tails in the line
shape associated with the noise in the dipole moment as well as to the intrinsic
quantum noise. 

In the model we study, we neglect the noise source causing
decoherence of the superconducting phase difference: voltage fluctuations. The
superconducting phase difference is fixed only for ideal voltage bias. In case
of nonideal bias, the voltage fluctuations will drive the phase evolution and
cause the drift of the superconducting phase difference. The corresponding
coherence time can
be estimated using the voltage correlation function for a junction with
impedance $Z$
	\begin{align}
		\langle V(t)V(t')\rangle = \delta(t-t')k_B T Z
	\end{align}
with $k_B T$ the thermal energy for temperature $T$. With this the average
value of the phase is given by
	\begin{align}
		\langle e^{i\phi(t)}\rangle =
			e^{-\frac{1}{2}\llangle\phi(t)\phi(t')\rrangle} =
			e^{-4\frac{e^2}{\hbar^2}k_B T Z t}
	\end{align}
where the second Josephson relation and the voltage correlation function
were used for the final expression. From this we find a coherence time of
$\tau_\text{dec,SC}= \hbar^2/4e^2k_BTZ$. The coherence time can thus be
prolonged by reducing temperature or impedance of the nanostructure.
Alternatively one can increase the number of Josephson junctions to reduce the
noise. The latter approach is used in metrology to realize the Josephson voltage
standard. A voltage drift of a few nV per hour can be
realized\cite{voltagestandard} in that way, leading to coherence times on the
order of $10^{12}$ s. Hence the voltage fluctuations may be disregarded
when studying the decoherence in the device.

This article is organized as follows. In Sec.~\ref{sec:I} we derive the
governing equations for a broader class of devices, characterized by a large
number of emitters. We start by briefly treating the model of the HJL, as
described in Ref.~\onlinecite{godschalk}, and subsequently argue that the
resulting
equations can be generalized to represent a wide range of devices. In
Sec.~\ref{sec:II} it is shown that lasing with a phase lock occurs for these
general equations in certain parameter regimes. There are distinguished regimes,
separated by first and second order phase transitions. In Sec.~\ref{sec:III}
conditions are derived to determine the parameter regime where the noise can be
considered in linear approximation. The frequency dependence of the noise,
when in a stationary lasing state, is calculated. The noise in linear
approximation cannot cause the decoherence of
the laser owing to the phase lock. However, occasional large fluctuations will
cause decoherence. These are treated in Sec.~\ref{sec:IV}, for which the
formalism of optimal paths is used. Finally, we conclude in Sec.~\ref{sec:V}.

\section{Model}\label{sec:I}

The purpose of this section is to derive a rather general but simple model
describing a novel class of lasing devices, those are driven by superconducting
leads biased at voltage $V$, have the optical resonance frequency at half
the Josephson generation frequency, $2eV/\hbar$, and contain many quantum
states capable of emitting light. The purpose
of the general approach is to capture the essential properties of this class of
devices, without involving specific microscopic details of
the used model. In this section we start by briefly repeating the essentials of
the model for the HJL at microscopic level and extend it to contain a large
number of effective emitters, yet at the microscopic level. We formulate the
generic features of the model which are not specific to the details of the
microscopic description. Finally, we present the general but simple model
where all these generic features are incorporated. We formulate the model in
terms of a Fokker-Planck equation for three variables.

\subsection{The HJL with a single emitter}
The HJL is described in Ref.~\onlinecite{godschalk} as an optical resonator
driven by a single quantum emitter. The emitter is a double quantum dot in a
{\it p-n} semiconductor nanowire, capable of containing up to two electrons and
two heavy holes. The energy of the corresponding discrete quantum states is
determined by orbital and Coulomb energies and is represented by Hamiltonian
$\hat{H}_\rs{QD}$. Owing to the superconducting proximity effect, a gap is
induced at the quantum dots for both electrons and holes, represented by
Hamiltonian
	\begin{align}\label{eq:ham_SC}
		\hat{H}_\rs{SC} =
\Delta_e^*\hat{c}_\uparrow \hat{c}_\downarrow
	+ \Delta_h\hat{h}_\uparrow \hat{h}_\downarrow + \text{H.c.}
	\end{align}
where the $\Delta_{e,h}$ are the proximity induced pair potentials, with the
phases $\phi_{e,h}$ retaining the values of the corresponding superconducting
leads. Owing to gauge invariance, only the superconducting phase difference,
$\phi_\Delta=\phi_e-\phi_h$ will have a physical significance.
Interaction of the double quantum dot with the resonator mode occurs according
to
	\begin{align}\label{eq:ham_photon}
		\hat{H}_\rs{ph} = \hbar\omega \hat{b}^\dagger \hat{b} 
		+ G (\hat{b}^\dagger \hat{x} + \hat{b}\hat{x}^\dagger)
	\end{align}
where $\omega = \omega_0-eV/\hbar$ is the frequency detuning from resonance,
with $\omega_0$ the resonance frequency of the optical resonator and $V$ the
bias over the device. The operator $\hat{x}=\hat{h}_\downarrow \hat{c}_\uparrow
+ \hat{h}_\uparrow \hat{c}_\downarrow$, taken in rotating frame, represents
the electron hole recombination required for the photon emission. It can be seen
as the dipole moment operator that drives the resonator mode. Coefficient $G$
is the coupling strength of the dipole moment to the resonator mode. It is
given by the dipole strength owing to quantum fluctuations of the electric field
in the resonator mode. The interaction Hamiltonian is assumed to be perturbative
as to allow for a large photon number before nonlinearities become manifest,
which will induce saturation of the dipole moment.

The dynamics of the resonator mode can be treated in a semiclassical fashion,
because a large number of photons will occupy the mode. In view of
this, the $\hat{b}^\dagger$, $\hat{b}$ and $\hat{b}^\dagger\hat{b}$ will be
replaced by their respective expectation values in the photon-dependent part of
the full Hamiltonian. A scaled quantity is defined, $\lambda\equiv G\langle
\hat{b} \rangle$. Within this approximation, the expectation value of the
dipole moment, $x_m(\lambda)\equiv\langle m|\hat{x} |m\rangle $,
is found by taking the derivative of the spectrum $E_m(\lambda)$ to $\lambda^*$,
with $E_m(\lambda)$ and eigenstates $|m\rangle$ associated with the full
Hamiltonian, $\hat{H}_\rs{QD}+\hat{H}_\rs{SC}+\hat{H}_\rs{ph}$. Hence, the
value of the dipole moment, $x_m$, depends on the particular eigenstate of the
double quantum dot. On the one hand, the dipole moment depends on the
optical field, $\lambda$, but on the other hand, the
evolution of $\lambda$ is determined by the dipole moment. This leads to a
set of self-consistent equations
	\begin{align}\label{eq:sceqn}
		\dot\lambda = -\left(i\omega +
\frac{\Gamma}{2}\right)\lambda
		- i \frac{G^2}{\hbar} x_m(\lambda), \quad
x_m = \frac{\partial E_m}{\partial \lambda^*}.
	\end{align}
Stationary lasing is found when $\dot\lambda=0$ for finite values of $\lambda$.
This state must be stable against small perturbations.

To get lasing, the rate at which photons are created in the resonator mode,
$2(G^2/\hbar)\text{Im}[x_m(\lambda)/\lambda]$, needs to exceed the
cavity escape rate, $\Gamma$. In common lasers, the imaginary part of the
susceptibility, Im$[x_m(\lambda)/\lambda]$, results from pumping to excited
states. Usually, to achieve lasing, the pumping must be
sufficiently intense to invert the population. For the HJL, it is the
superconductivity that creates the required complex susceptibility so that the
population inversion is not necessary. In fact, the ability of the
superconductors to absorb
or supply Cooper pairs undermines the notions of groundstate and excited state
in the quantum dots. For instance, the
induced pair potentials enable the possibility of transitions from lower to
{\it higher} energy states, while {\it emitting} a photon, with the required
energy provided
by the bias. This feature allows for cycles of emission or absorption of
photons, in order to (de)populate the resonator mode. 

Decoherence in the HJL with a single emitter is caused by spontaneous
switchings between quantum dot eigenstates. Owing to the finite cavity
linewidth, emission of off-resonant photons occurs inducing these switchings
They occur at a slow rate, $\Gamma_\rs{SW}\simeq\Gamma/n\ll\Gamma$, and
therefore introduce extra dynamics in the HJL described by a master equation.
This is in addition to the already present dynamics of Eq.~\eqref{eq:sceqn}. It
was found in Ref.~\onlinecite{godschalk} that even for an optimal choice of
parameters, not every quantum dot eigenstate couples to the resonator mode
strongly enough to support lasing. The ones that do support lasing will yield
different stationary values $\lambda_s$. Since only some states support
lasing, the optical field may even extinguish after a switch, to turn on again
after another one. Hence, the switchings change $\lambda_s$, causing large
intensity and phase fluctuations. Most notably, for every lasing state
$\lambda_s$ there is another one at $-\lambda_s$. When going from some
lasing state to a nonlasing state and returning to the same lasing state again,
the optical field can go to both $\pm\lambda_s$ with equal
probability. Hence the switchings are responsible for decoherence. The
coherence time is determined by the switching rate, $\tau_\text{dec}\simeq
n/\Gamma$, and is similar to that of the common laser.\cite{scullylamb}

\subsection{The HJL with many emitters}
Even though the HJL, as described above, has a rather specific design, the
framework provided by Eqs.~\eqref{eq:ham_SC}-\eqref{eq:sceqn} is
quite general. It applies to a class of devices where a biased Josephson
junction contains a structure of which the eigenstates couple to both
superconducting leads and which can emit light by electron hole recombination.
The light is emitted into an optical resonator. When the structure in the
Josephson junction is large enough it can be divided into $N\gg1$ blocks. Every
block represents an effective quantum emitter, which is a quantum system with a
number of discrete eigenstates that can emit light. Every emitter must couple to
both superconducting leads and must emit light into the resonator mode. In
principle, the emitters interact with each other, but we may model them to be
independent. As an example of such a structure, one might consider an ensemble
of {\it p-n} semiconductor nanowires with a double quantum dot, the nanowire
that was used in the previous section, or one might consider a two-dimensional
electron gas between the superconducting
electrodes.\cite{S-2DEG-S}

Increasing the number of emitters in the Josephson junction to enhance the
lasing is a method reminiscent of a line of research where superconducting
lasers are made based on stacked Josephson junctions.\cite{stackedjj} There,
the Josephson junctions are connected in series so that cascade amplification of
the radiation occurs. The setup of the HJL is different though, because the
emitters can be regarded as Josephson junctions connected in parallel rather 
than in series.

The dipole moment corresponding to the structure in the Josephson junction is
determined by its energy, which is taken in a rotating frame of reference. When
the structure contains $N$ emitters the total energy is given by the sum of the
energies of the emitters. Assuming that emitter $i$ is in a state with energy
$E_i$, the total energy is simply $E=\sum_i E_i$. This yields a value of the
total dipole moment given by $D=\partial E/\partial\lambda^*$, which should be
inserted in Eq.~\eqref{eq:sceqn} to look for solutions that support lasing,
$\lambda_s$.

Every emitter is subjected to spontaneous switchings that change its eigenstate.
Therefore, at time scales longer than the switching times it is not the total
energy (dipole moment) that is relevant but the average total energy (dipole
moment). The switching dynamics defines a master equation for each of the
emitters, where the probability $p_m^i$ of emitter $i$ to be in
state $m$ evolves as
	\begin{align}\label{eq:masterequation}
		\frac{d p_m^i}{dt} = \sum_{n\ne m}[ W_{m;n}^i\delta_{nm} -
W_{n;m}^i ]p_m^i
	\end{align}
with $W_{m;n}^i$ the transition rate from state $|n\rangle_i\to |m\rangle_i$ in
emitter $i$. The quantity in brackets $[\cdots]$ defines a transition matrix,
which necessarily has a null vector to ensure the conservation of probability.
The null vector gives the stationary occupation probabilities, $(p_m^i)_s$. With
these, the average total energy is given by $\langle E\rangle =
\sum_{i,m} (p_m^i)_s E_m^i$, where $E_m^i$ is the energy of state $m$ in emitter
$i$. Because of the coupling to the resonator mode, the energy depends on
$\lambda$ and the total average dipole moment is $\langle D\rangle =\partial
\langle E\rangle/\partial\lambda^*$. Now this quantity determines the lasing
solutions of Eq.~\eqref{eq:sceqn}.

\subsection{The average value of the dipole moment}
Let us discuss a complication that might easily arise when considering this
model or designing a device. In our approach we consider spontaneous
switchings only to be a result of off-resonant photon emission. In the limit
of a large number of emitters the associated transition rate matrix,
Eq.~\eqref{eq:masterequation}, turns out to be approximately symmetric, with
$W_{m;n}^i\approx W_{n;m}^i$.
The transition rates $W_{m;n}^i$ are calculated using Fermi's golden rule. They
are, up to some constants, given by the product of a matrix element squared,
$|\langle m(\lambda)|\hat{H}_\rs{ph}| n(\lambda) \rangle|^2$, for a given
emitter $i$, and a Lorentzian shaped density of states, $\rho(E_{mn}^i) =
(1/2\pi)\hbar\Gamma/[(\hbar\Gamma)^2 + (E_{mn}^i)^2]$, with $E_{mn}^i$ the
energy difference between the eigenstates $|m(\lambda)\rangle_i$ and
$|n(\lambda)\rangle_i$. We note that the photon number
generically remains the same during a transition; $\lambda$ is constant. Only
after the transition may relaxation to the new stationary state occur. For a
large number of emitters, however, a single switching is not enough to change
the
electrical field in the mode significantly. The stationary state $\lambda_s$
remains almost the same after the transition and therefore also the eigenstates
and energies remain approximately the same. Consequently, the transition rates
also remain the same. Since the energies $E_{mn}^i$ have a large
mismatch with the resonance frequency (the emitted photons are far off-resonant)
they do not depend significantly on detuning, $\omega$, and we have
$E_{mn}^i=E_{nm}^i\gg\hbar\Gamma$, for fixed $\lambda$. The transition rate
from state $|m(\lambda)\rangle_i$ to $|n(\lambda)\rangle_i$ is therefore equal
to the transition rate of the reversed process. Hence, because the transition
rates remain approximately the same after a transition, the transition rate
matrix is approximately symmetric.

Innocent as it may seem, this feature of the transition matrix poses a problem
for our device, because as a result the average value of the dipole moment,
$\langle D\rangle$, must be approximately zero. The null vector for a symmetric
transition matrix must be a vector with all entries
equal. Therefore, in the stationary solution to the master equations, all the
states in an effective emitter $i$ have the same probability $p_s^i$ to be
occupied. The total average dipole moment is then given by $\langle D\rangle =
\sum_{i} p_s^i \sum_m x_m^i$. However, for the summation over $m$ we find
	\begin{align}\label{eq:zerodipole}
		\sum_m x_m^i(\lambda) = 
			\frac{\partial}{\partial\lambda^*} \sum_m E_m^i =
			\frac{\partial}{\partial\lambda^*} \text{Tr}_i[\hat{H}] = 0,
	\end{align}
where Tr$_i$ is the trace over the space of states of emitter $i$. The
equality to zero follows because Tr$_i[\hat{H}]$ is independent of $\lambda^*$.
Consequently $\langle D\rangle = 0$. We note that this is generic for
any kind of system governed by our Hamiltonian and where the
relevant states defining the dipoles, are equally populated. Naively one would
expect that a large number of emitters would result in a large number of
photons in the resonator mode. We see that this is not so. On the contrary, the
total dipole moment could be zero and no photons are present.

We need to adjust the model used in Ref.~\onlinecite{godschalk} to get a nonzero
average dipole moment. This is done by making the transition rate matrix
asymmetric. Then the occupation probabilities for the states of an emitter
will not be equal anymore, resulting in an imbalance in populations of states.
There is no {\it a priori} reason to favor, for instance, the state with higher
energy
over the state with lower energy, for higher occupation probability. The notion
of higher or lower energy of states has become obscured by the superconducting
pair potentials. Hence, the created imbalance is not the same as a population
inversion. Importantly, recombination of electron-hole pairs will still be
characterized by the phase lock, because they are still coherent with the
superconductors. Despite introducing incoherent transition rates, the special
characteristics of the device are thus maintained.

An asymmetry in the transition rate matrix can come about in various ways. 
The simplest way is to take into account the photon emission into the
environment, rather than to the resonant mode. If the rate of this incoherent
process is comparable with the emission rate to the resonant mode, the average
dipole moment is of the order of its maximum value per emitter.
In a realistic system we expect relaxation transitions between the
emitter states that do not involve photon emission. These relaxation channels
will generally yield asymmetric transition rate matrices. One can
also engineer transition rates to get a sufficiently asymmetric
transition matrix. A straightforward approach is to change the rates that are
already present by changing the electromagnetic environment. A switching to a
state with higher (lower) energy is accompanied by emission of an off-resonant
photon, with energy $eV-E_{nm}$ ($eV+E_{nm}$), where $E_{nm}\gg \Gamma$. By
changing the density of states in favor of one of these two, the spontaneous
emission rate for the corresponding process increases with respect to the other
one, thus creating the desired imbalance in transition rates. This might be done
by inserting the device in a low quality factor resonator with resonance
frequency about either of the frequencies. The low quality factor will ensure
the rapid escape of the photon, before it is reabsorbed again. Alternatively,
the electromagnetic environment can be changed by shining with a red or blue
detuned laser on the device, favoring transitions to lower and higher energy
eigenstates respectively, proportional to the laser intensity. This resembles
the technique of resolved sideband cooling/heating.\cite{diedrich}
Another possibility is to add new transition rates to the system by adding
nonsuperconducting leads to the device and
directly pump electrons and/or holes into the optically active region. This,
however, leads to big challenges in the technical device design.

\subsection{Phase dependence of the energy}
We have discussed the general features of possible microscopic realizations of
a HJL with a large number of emitters. Let us construct a model that does not
rely on microscopic details. The relevant quantity to deal with is the
$\lambda$-dependent part of the total energy of the device, which determines the
dipole moment. We show that the energy has a generic dependence on the phase
difference between the superconducting leads. 

Generally, we note that both common lasers and superconductors are systems
characterized by a spontaneously broken $U(1)$-symmetry, a phase symmetry. As
such they are characterized by an order parameter which has a magnitude and a
phase. Any actual realization of the order parameter bears no apparent
connection to the original unbroken symmetry.
However, as a remnant of the symmetry, every value of the phase is equally
probable and corresponding to states with equal energy. States with different
phases are degenerate so that the energy of the system is independent of the
phase.
 
Our model combines both of these systems, lasers and superconductors, which
leads
to a phase lock between the optical phase and the superconducting phase
difference.
Indeed, owing to the interaction between these two systems a phase dependence
must enter the energy of the system so that not both of the spontaneously broken
symmetries can be maintained. The emission of a single photon is caused by the
recombination of an electron-hole pair, with the latter carrying the phase of
their respective condensates. The phase of the terms in the relevant interaction
Hamiltonian, Eq.~\eqref{eq:ham_photon}, is therefore given by
$\phi_\lambda-\phi_\Delta/2$, with $\phi_\lambda$ the optical phase and
$\phi_\Delta$ the superconducting phase difference. In fact, the full
interaction depends on the cosine of this phase combination, leading to a
similar dependence in the energy. As a
consequence an energy minimum is found for a specific value of
$\phi_\lambda-\phi_\Delta/2$, which
is a manifestation of the phase lock between optical phase and
superconducting phase difference. A further inspection of the phase dependence
of the energy shows that there must, in fact, be two equivalent values of the
phases. Changing $\phi_\Delta\to\phi_\Delta+2\pi$ leaves the
system invariant since the superconducting phase difference has not changed, but
the phase in the interaction has changed to $\phi_\lambda -\pi-\phi_\Delta/2$,
so we conclude that $E(\phi_\lambda)=E(\phi_\lambda-\pi)$. Hence, the original
spontaneously broken $U(1)$ symmetries of the
Hamiltonian have been reduced to a single one and a spontaneously broken
symmetry under rotations of $\pi$ of the optical phase. The
remaining $U(1)$ symmetry comprises invariance under a simultaneous
change in optical phase and superconducting phase difference such that the
quantity $\phi_\lambda-\phi_\Delta/2$ is unchanged. 
We stress that this reasoning applies to any system that combines two
order parameters with spontaneously broken $U(1)$ symmetries, with an
interaction as in Eq.~\eqref{eq:ham_photon}. Any such system manifests a
phase lock.

\subsection{Energy}
To find lasing in the general model, we need to have an expression for
the energy of the device. This allows us to extract essential physics
from the model.

Generally, the energy should depend on the value of light field in the
resonator, $b\equiv \langle\hat b\rangle$, and the pair potentials,
$\Delta_e,\Delta_h$, in the superconductors: $E(b,\Delta_e,\Delta_h)$. We
use parameter limits where the four variables are relatively small,
enabling us to use a Taylor expansion for the energy. This limit corresponds to
a weak drive of the resonator mode. To define ``small'' we go
back to the model where the dipole medium consists of many emitters. We can
associate an energy scale, $E_s$, to each of these emitters. The induced pair
potentials should be small compared to this scale, $\Delta_{e,h}\ll E_s$. In a
realistic device this can be engineered by making suitable (tunnel) barriers
between the superconductors and emitters. For the second limit, it is
reasonable to take the interaction between dipole and resonator mode in the weak
coupling limit. Hence the coupling constant $G$ of the interaction Hamiltonian,
Eq.~\ref{eq:ham_photon}, is much smaller than $E_s$. Therefore, this
interaction perturbs the dipole energy by a value of the order of $E_s$ only
when a large number of photons, $n_0=|b_0|^2$, occupies the resonator mode. We
want to make sure that the stationary value of photon number, $n_s$, is always
much smaller than $n_0$, so that the use of an expansion of the energy in terms
of $b$ is always justified.

When writing the Taylor expansion of the energy,
$E(b,\Delta_e,\Delta_h)$, only a limited number of terms are allowed
because of the specific phase dependence that is required. As was
discussed in the previous section, the energy should depend on the phase
combination, $\varphi =2\phi_\lambda-(\phi_e-\phi_h)$. Therefore, the lowest
order phase-dependent term in the energy must be proportional to
$\Delta_e^*\Delta_h b^2 + $complex conjugate. This is a fourth-order term
in the Taylor expansion. All other terms of equal or lower order are
proportional to powers of $|b|^2$ or $|\Delta_{e,h}|^2$. Writing only the
terms with immediate relevance, the energy is, up to fourth-order in Taylor
expansion
	\begin{align}\label{eq:energy}
		E = E_0 +\Omega'|b|^2 + \frac{1}{2}\Omega''|b|^4 +
			\frac{A}{2} ({b^*}^2+ b^2)
	\end{align}
In this expression the phase of the induced pair potentials was absorbed into
the optical phase, so it is redefined as $\phi_\lambda \to
\varphi/2 = \phi_\lambda-\phi_\Delta/2$. Furthermore, we absorbed the induced
pair potentials in the prefactor $A = C |\Delta_e\Delta_h|$, with $C$ a
constant. The primes in $\Omega'$ and $\Omega''$ refer, respectively, to first
and
second derivatives at $|b|^2 = 0$ of the function $\Omega(|b|^2)$, which
is independent of the induced pair potentials. The lowest order term, $E_0$, is
a sum of $\Omega(0)$ and some other terms depending on $|\Delta_{e,h}|^2$ but
not on $b$ and $b^*$. These terms will therefore not be relevant for the dipole
moment. There are also terms proportional to $|\Delta_{e,h}|^2|b|^2$, but these
are small compared to $\Omega'$, because they are of higher order in the Taylor
expansion, and can safely be neglected. It is necessary to keep the $|b|^4$ term
in the expansion, since this nonlinear term stabilizes a possible lasing
instability. Without this term the light field would grow exponentially under
unstable conditions. When calculating the dipole moment later, we will see
that $\Omega'$ adds up to the detuning from resonance. Hence, the value of
$\Omega'$ can implicitly be changed by changing the detuning. In fact, to get
lasing the detuning should be such that $\Omega'$ effectively becomes small, of
the order of $A$ or smaller, or goes to zero. Because the detuning $|\omega|\ll
eV/\hbar$, this also sets the scale $|\Omega'|\ll eV/\hbar$. We stress that this
is an {\it effective} change of $\Omega'$. In the expansion of
Eq.~\eqref{eq:energy} the $\Omega'$ term is still considered to be larger than
the fourth-order terms. With effectively $\Omega'\lesssim A$, the $|b|^2$ term
is of the same order as, or smaller than, the $|b|^4$ term so the latter should
be maintained in the expression for the energy. 
With the above expression for the energy we have managed to
separate the dependence of the energy on $b$ and $\Delta_{e,h}$ from the
microscopic details of the driving medium. The latter are all contained in the
prefactors of the Taylor expansion.

Substituting Eq.~\eqref{eq:energy} into Eq.~\eqref{eq:sceqn}, we find an
explicit, general self-consistency equation. Noting that $\lambda =
G\langle b\rangle$ and $\partial E/\partial b^* = G\langle d\rangle$, we have
	\begin{align}\label{eq:sceqngen}
		\dot b = -\left(i \tilde\omega + \frac{\Gamma}{2}\right) b 
						- i (A b^*+\Omega''|b|^2b),
	\end{align}
where $\hbar=1$ and $\tilde\omega \equiv \omega + \Omega'$ is the redefined
detuning. This equation describes the deterministic evolution of the electric
field, $b$, in the resonator mode. We note that an equivalent complex conjugate
equation exists.

\subsection{Fokker-Planck equation}
The energy was considered to be a fixed quantity in the above section,
corresponding to the average value of the dipole moment. To fully account for
all dynamics in the HJL with many emitters, we also have to consider noise.
This can be incorporated in a single Fokker-Planck equation that encompasses
two noise sources. The first noise source is the traditional quantum noise due
to photon emission from the mode.\cite{risken} The second noise source
describes the fluctuations of parameters in Eq.~\eqref{eq:energy}.

To understand the latter let us note that Eq.~\eqref{eq:energy} is valid not
only in average, but also for each specific configuration of the emitter
quantum states. The parameters $\Omega'$, $\Omega''$, and $A$ are configuration
dependent. Since the emitters undergo spontaneous transitions the parameters
fluctuate accordingly. Explicitly taking into account all the
details of the fluctuations would greatly complicate the model. Instead, we
employ the central limit theorem stating that any ensemble of equal
distributions can be described as a Gaussian distribution when the number of
distributions goes to infinity. Therefore, since we have a large number of
emitters, as described above, we may regard the parameters as
stochastical quantities with Gaussian distributions.

In principle, the noise in all parameters is expected to be of
the same order of magnitude relative to their average values. However, since
$\Omega'$ is of lower order in the Taylor expansion of Eq.~\eqref{eq:energy},
its fluctuations are more important than that of $\Omega''$ and $A$. We
therefore only consider a single
noise source by taking $\Omega''$ and $A$ to be equal to their average values,
while $\Omega'\to\Omega'_e+\tilde\Omega$, with $\Omega'_e$ being the average
value and $\tilde\Omega$ being the deviation. The latter appears in
Eq.~\eqref{eq:sceqngen} as an addition to the detuning. The noise can therefore
be seen as fluctuations in detuning or in the resonant frequency. The
corresponding variance, $\langle \tilde\Omega^2\rangle$, is an important
parameter of our general model. We shall take into account that emitters switch
on a certain time scale that determines autocorrelations of these detuning
fluctuations. The autocorrelation function is thus given by\cite{vankampen}
	\begin{align}\label{eq:dipoleautocorr}
	   \llangle\tilde\Omega(t)\tilde\Omega(t+\tau)\rrangle = 
			\langle\tilde\Omega^2\rangle e^{-\gamma|\tau|},
	\end{align}
where $\gamma$ is the inverse autocorrelation time of the emitter ensemble

It is possible to capture all relevant dynamics compactly in a single nonlinear
Fokker-Planck equation. This is an equation for the probability distribution
function $P(b,\tilde\Omega,t)$ of three real values, two of them are represented
in the optical field and the third in the detuning deviation. The drift and
diffusion terms in the equation respectively
describe deterministic and noise driven, indeterministic
processes.\cite{vankampen} The Fokker-Planck equation for our device
encompasses
two Gaussian noise sources, describing fluctuations in three quantities.
First, there is the dynamics of $b$ which is subjected to a noise source,
$\tilde b$, describing the fundamental fluctuations in $b$. This is the
quantum noise associated with the discrete changes of photon number in the
resonator mode. It satisfies, $\langle\tilde b(t)\rangle=0$ and $\langle
\tilde b(t) \tilde b^*(t')\rangle = \Gamma\delta(t-t')$. Second, as discussed,
the noise in $\Omega'$ is equivalent to detuning fluctuations. This is
represented by the quantity $\tilde\Omega$, with average
$\langle\tilde\Omega\rangle=0$ and variance $\langle \tilde\Omega^2\rangle$.
The Fokker-Planck equation reads
\begin{widetext}
	\begin{align}\label{eq:FPE}
		\frac{\partial P(b, \tilde\Omega,t)}{\partial t} = 
		2\text{Re}\left[\frac{\partial}{\partial b}\left\{ 
\left[i(\tilde\omega +\Omega''|b|^2+\tilde\Omega)+  
		\frac{\Gamma}{2}\right]b P + i A b^* P \right\} \right]
		+ \frac{\Gamma}{2}\frac{\partial^2 P}{\partial b\partial
b^*} +
		\gamma \left\{ \frac{\partial}{\partial \tilde\Omega}
\tilde\Omega P + \langle {\tilde\Omega}^2
		\rangle
		\frac{\partial^2 P}{\partial {\tilde\Omega}^2}   \right\},
	\end{align}
\end{widetext}
The single derivative terms on the right-hand side of Eq.~\eqref{eq:FPE} are
the drift terms. Evaluating $\langle \dot b\rangle =
\int db\, b \dot P$ by using Eq.~\eqref{eq:FPE} will yield the self-consistency
equation~\eqref{eq:sceqngen} for $\langle b\rangle$. We remark that in this
context the brackets, $\langle\cdots\rangle$, denote the average over the
classical distribution $P$, and {\it not} the expectation value of the
operator $\hat b$. The diffusion is described by the second derivative terms on
the right-hand side of Eq.~\eqref{eq:FPE}.

Let us note that all the parameters we use have the dimension of frequency. It
is convenient to make them dimensionless, measuring them in relative units of
$\Gamma/2$. This amounts to setting $\Gamma\to2$ in Eq.~\eqref{eq:FPE}. 
For clarity, we rescale the detuning deviation in Eq.~\eqref{eq:FPE} to
appropriate units, by taking $\tilde \chi \equiv \tilde\Omega/\sqrt{A^2 -1}$,
which is the scale of the frequency domain where lasing is possible, as we show
in Sec.~\ref{sec:II}. Additionally we have
$\langle\tilde\chi^2\rangle\equiv \langle\tilde\Omega^2\rangle / (A^2-1)$. As a
consequence, the term in Eq.~\eqref{eq:FPE} proportional to $i\tilde\Omega b P$
changes to $i\sqrt{A^2-1}\ \tilde\chi b P$. Other terms retain their form.

We conclude this section by stressing that this Fokker-Planck equation applies
to a whole class of lasers, where a biased Josephson
junction contains a structure of which the eigenstates couple to both
superconducting leads and which can emit light by electron-hole recombination
into the resonator mode.

\section{Lasing}\label{sec:II}

In the previous section we have derived a model for a class of lasers that is
driven by superconductivity and that exhibits a phase lock.
Here we show that lasing with a phase lock is indeed possible and we give
expressions for the lasing threshold and the number of photons in the resonator
mode. We find three different regimes, with a different number of
stationary lasing states. 

\begin{figure}
  \centering
  \includegraphics{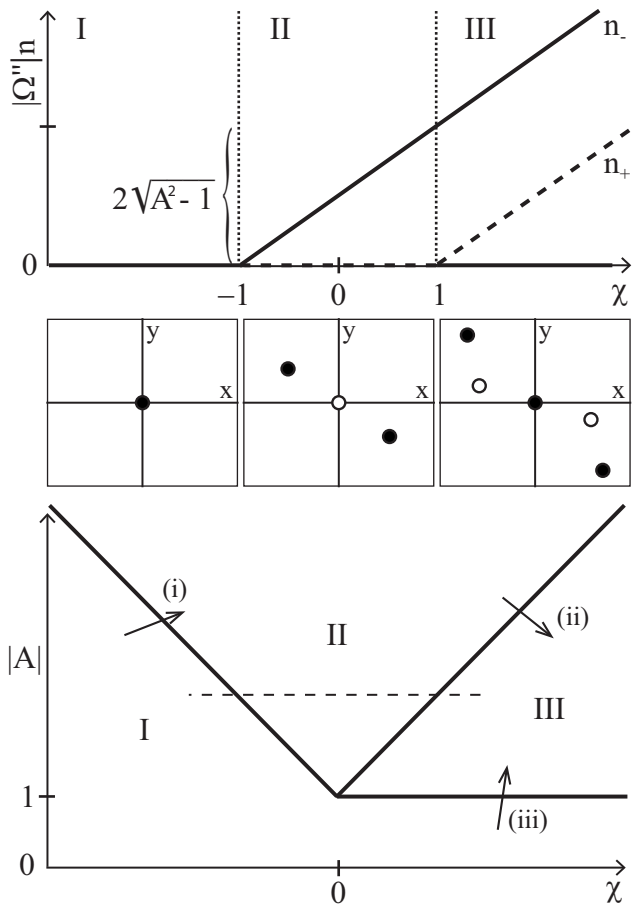}
  \caption{%
  Lasing in the device. The top plot shows the number of photons,
$n$, for the stationary solutions of Eq.~\eqref{eq:statsol}, as a function of
the rescaled detuning $\chi$, with $\Omega''< 0$. Thick solid (dashed) lines
represent
stable (unstable) solutions, labeled by $n_\pm$. Three parameter regimes, (I,
II and III), are found based on the number of stationary solutions. The
vertical dotted lines represent boundaries between these regimes. Below this
plot is a panel that illustrates the existing stationary solutions in the
$x$-$y$ plane, for each regime. Solutions with $n>0$
occur twofold with opposite sign. Solid (empty) circles represent stable
(unstable) stationary states.
The bottom panel shows the dependence of the regimes on the parameters $|A|$
and $\chi$. The diagonal lines are the lasing thresholds $\chi_\pm$ as discussed
in the main text. The line between regimes II and III is the one where
$n_\pm$ becomes real-valued. The plot of the upper figure is taken along the
dashed line.
When crossing the boundary along arrow(s) (i) and (ii) [(iii)], the system
undergoes a second [first] order phase transition. 
  }\label{fig:lasing}
\end{figure}

A full description of the dynamics of lasing would require solving Fokker-Planck
equation~\eqref{eq:FPE}, but this is difficult, if not impossible, even for a
relatively simple system such as ours. Instead, we first search for
stationary lasing solutions to the deterministic equation of motion,
Eq.~\eqref{eq:sceqngen}, thus disregarding noise. Once
these solutions are found we study the effects of noise when near a stable
stationary lasing state. This is the topic of the subsequent two sections.

To find the desired stationary solutions to Eq.~\eqref{eq:sceqngen}, with $\dot
b=0$, we rewrite them as real equations. Defining $b\equiv x+iy$ leads to a set
of equations,
	\begin{align}\label{eq:statsol}
		x = (\tilde\omega -A +  \Omega'' n)y, \quad
		y = -(\tilde\omega +A +  \Omega'' n)x,
	\end{align}
where $n=x^2+y^2$ is the photon number in the resonator mode. These equations
can be readily solved to find an expression for $n$ and for the phase of $b$,
which was defined as $\varphi/2=\phi_\lambda-\phi_\Delta/2$, 
	\begin{align}\label{eq:scsoln}
		\begin{split}
		n_\pm &= \frac{\sqrt{A^2-1}}{\Omega''}\left[ \pm 1 -
				\chi \right], \\
		\tan\left(\frac{\varphi_\pm}{2}\right) &= -A\mp \sqrt{A^2-1}, 
		\end{split}
	\end{align}
where we conveniently rescaled the detuning, $\tilde\omega=\chi \sqrt{A^2-1}$.
Hence, we find stationary lasing 
solutions. The fixed value of $\varphi_\pm$ implies that $\phi_\lambda$ 
and $\phi_\Delta$ are locked. Note that the
equations are invariant under the transformation $x\to-x$ and $y\to-y$, implying
the occurrence of two solutions which differ in phase by $\pi$,
as predicted in the previous section. Indeed, Eqs.~\eqref{eq:scsoln} are
invariant under the transformation $\varphi_\pm/2\to \varphi_\pm/2+\pi$. Hence
$n_\pm$ and $\varphi_\pm$ represent four stationary solutions. Besides these,
there is another stationary solution, being $x=y=0$, delivering a total of five
possible stationary solutions. We remark that $\varphi_\pm$ only depends on $A$
and not on $\chi$ and $\Omega''$. Furthermore, a large photon
number can be achieved when $\Omega''\ll A$.

Studying the stationary solutions of Eq.~\eqref{eq:scsoln} we find three
different regimes that differ in the number of physical solutions. Since
$n_\pm$ represents the number of photons in the resonator mode, it must
be either a real and positive quantity or it must be zero. As a result we note
that lasing solutions, with $n_\pm>0$, can only be achieved when $A^2>1$.
Then, depending on $\chi$, three different regimes exist
(Fig.~\ref{fig:lasing}). We choose $\Omega''<0$ but similar regimes can be
found when $\Omega''>0$. For $\chi<0$ and $|\chi|>1$, both $n_\pm<0$ so
that the only physical solution is $n=0$. We call the parameter regime with a
single solution regime I. Regime II is given by the condition that $|\chi|<1$,
where $n_+<0$ and $n_->0$. Including $n=0$ there are now three physical
solutions. The third regime is when $\chi>0$ and $|\chi|>1$,
where both $n_\pm>0$, corresponding to a total of five physical solutions.

Even though there can be multiple physical solutions, not all of them will
support stationary lasing, because some will be unstable against small
perturbations. To find the condition for stability we expand
Eq.~\eqref{eq:sceqngen}, in terms of $x$ and $y$, up to first order about a
stationary value. This yields
	\begin{align}\label{eq:stability}
		\frac{d}{dt}\begin{pmatrix} \delta x \\ \delta y
\end{pmatrix} =
			\begin{pmatrix} 
				-1+2\Omega''xy & \frac{ x}{y}+2\Omega''y^2\\
				\frac{ y}{x}-2\Omega''x^2 & -1-2\Omega''xy
			\end{pmatrix}
		\begin{pmatrix} \delta x \\ \delta y \end{pmatrix},
	\end{align}
where $x$ and $y$ are at the stationary values, $\delta x$ and $\delta y$ are
the respective deviations from the stationary values and Eq.~\eqref{eq:statsol}
was used to rewrite some terms. When an eigenvalue of the matrix is negative
(positive), a perturbation in the direction of the corresponding eigenvector
decays (grows) exponentially to (from) the stationary  value of $x$ and $y$. The
case where both eigenvalues are negative corresponds to a stable stationary
lasing state. After a fluctuation in any direction, the system will relax to the
stationary state again. With one positive and one negative eigenvalue there is a
stable and an unstable direction against perturbations, thus defining a
saddle point. The case where both eigenvalues are positive cannot occur,
because the trace of the matrix is smaller than zero. The condition
of the stability therefore depends on the sign of the determinant of the matrix,
positive being the stable stationary state. Evaluating the determinant and again
using Eq.~\eqref{eq:statsol} we derive that the stability for a physical
solution $n_\pm>0$ is determined by sign$\left[4(A^2-1)(1\mp\chi)\right]$,
whereas the stability of the solution at $n=0$ is determined by sign$\left[(A^2
- 1)(1 - \chi^2)\right]$. 

To illustrate how the system relaxes to a stable stationary lasing state after a
small perturbation, we give the eigenvalues of the matrix in
Eq.~\eqref{eq:stability}. These are, for the case of $n_-$, $\eta_{1,2} =
-1\pm i\sqrt{4(A^2-1)(1+\chi)-1}$ and, for the case of $n_+$, $\eta_{1,2} =
-1\pm i\sqrt{4(A^2-1)(1-\chi)-1}$. Hence, when $A$ is sufficiently large, the
system
oscillates about the stationary point with a frequency $\pm\simeq 1/A$ and
relaxes to it in $\simeq A/2\pi$ rotations.

For $\Omega''<0$
the stability of the stationary solutions in the various regimes is as depicted
in Fig.~\ref{fig:lasing}. Regime II is the most relevant one for lasing as
the $n=0$ solution is unstable, whereas the $n_-$ solution is stable. In regime
III the $n_-$ and $n=0$ solutions are all stable, whereas the $n_+$ solutions
are unstable. When $\Omega''>0$, the qualitative picture is
the same. Quantitatively, the location of the regimes changes. In the phase
diagram of Fig.~\ref{fig:lasing}, the regimes I and III are interchanged so that
the plots for $\Omega''>0$ are mirror images of the plots for
$\Omega''<0$. Additionally, the $n_+$ ($n_-$) solution is now the stable
(unstable) one. 

The boundaries between the regimes can be seen as lasing thresholds because they
separate the lasing regimes from the nonlasing regime. The boundaries
between regimes I and II and between regimes II and III are lines in parameter
space where the determinant of the matrix of Eq.~\eqref{eq:stability}
is zero at $n=0$. They are given by $\chi_\pm=\pm1$. For negative
$\Omega''$ the value
$\chi_-$ ($\chi_+$) corresponds to the boundary between regimes I
and II (II and III). The boundary between regimes I and III is
determined by the condition $A=1$. At this boundary, the $n=0$ solution is
still stable, but two pairs of stable and unstable solutions arise at values
$n>0$. When gradually crossing the boundaries from I to II or II to III, the
system undergoes a second order phase transition [arrows (i) and (ii) in
Fig.~\ref{fig:lasing}]. Thereby the $n=0$ state becomes unstable. In contrast to
this, when crossing the boundary from regime I to III [arrow (iii) in
Fig.~\ref{fig:lasing}] the system will undergo a first-order phase transition,
because all physical stationary solutions will arise
directly at nonzero values, two stable and two unstable. 
We note that $A$ cannot be increased indefinitely as the pairing potentials must
remain small enough to satisfy the adopted approximations.

To conclude this section, we find that regime II is the only regime where
lasing occurs without an interruption caused by a large fluctuation. This is
because only there the $b=0$ solution is unstable. In contrast to this, in
regime III a large fluctuation can bring the device from a lasing state to the
stable nonlasing state at $n=0$ thereby interrupting the emission of light.
Regime II occurs in a small interval at large detuning, where
$\sqrt{A^2-1}|\chi|=|\tilde\omega|=|\omega+\Omega'|\simeq |A|\ll|\Omega'|$.
The number of photons in the resonator mode, being of the order of
$|A/\Omega''|$, is large when $|\Omega''|\ll|A|$. In the remainder of this work,
we assume we have chosen the parameters $A$ and $\chi$ such that
the system is in regime II. We take $\Omega''<0$ so that the $n_-$ solution
is the lasing solution.

\section{Noise and its frequency dependence}\label{sec:III}

Let us consider fluctuations in the device. It is important to note that because
of the phase lock the laser cannot lose coherence by a random drift of the
optical phase as in common lasers. On the contrary, after a small fluctuation a
damping force
returns the system to the stable stationary lasing state. The average value of
the optical phase is therefore retained.  As in the HJL,\cite{godschalk} the
only possibility to lose coherence is when the device switches to another stable
stationary lasing state. This requires a sufficiently large fluctuation in
photon
number or optical phase. We envision these large fluctuations to occur only
rarely, on exponentially long time scales, implying that the noise in $b$ must
be
linear, i.e. dominated by small fluctuations about the stationary lasing state.
In this section we investigate the HJL in the linear noise approximation. There
are two noise
sources causing intrinsic quantum fluctuations in the resonator mode and
fluctuations in detuning. Therefore, the steady state of the HJL can be
described by a multivariate Gaussian probability distribution, for which we
calculate the variances. Several parameter limits will be discussed concisely
for the photon number variance. We also find an expression for the corresponding
noise spectrum. The large fluctuations that cause decoherence of the laser are
the topic of the next section.

\begin{figure}
  \centering
  \includegraphics{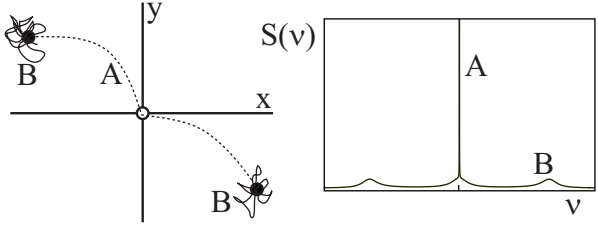}
  \caption{%
Noise and the noise spectrum in the HJL with many emitters. On the left-hand
side
is an illustration of fluctuations in the optical field, $b=x + i y$, where we
neglect for clarity the detuning noise on the $z$ axis. The value of $b$
evolves along the lines. The dashed line, indicated by $A$, represents a large
fluctuation, where $b$, starting in the vicinity of a stable lasing
point, crosses the unstable point to the other stable lasing point. This
process shall occur on exponentially long time scales,
$\tau_\rs{large}$ (see main text). The solid erratic lines, indicated with $B$,
represent small fluctuations in the vicinity of the stable points. These
fluctuations typically occur on a time scale $\tau_\rs{small}$ (see main text).
The two kinds of fluctuation contribute independently to the noise spectrum, of
which an illustration is shown on the right-hand side. The sharp peak,
indicated with $A$, corresponds to the large fluctuations. It is a
Lorentzian-shaped peak with exponentially small width. The small fluctuations
provide a
broadband background, indicated by $B$. Details on the peaks can be found in
the main text.
  }\label{fig:noise}
\end{figure}

We first describe an intuitive picture of the shape of the full noise spectrum,
containing the spectrum of both large and small fluctuations.
A sketch of this is given in Fig.~\ref{fig:noise}. The large
fluctuations occur at exponentially long time scales, $\tau_\rs{large}$,
satisfying $\ln(\tau_\rs{large}/\tau_\rs{small})\gg1$. Here, as we see
later in this section, the small fluctuations occur at much shorter time scales,
determined by the largest of the response time of the resonator mode,
$\simeq1/\Gamma$, or the autocorrelation time of the detuning noise,
$\simeq1/\gamma$, yielding $\tau_\rs{small}=\text{max}(1/\Gamma,1/\gamma)$.
Because of this separation of time scales, both kinds of fluctuation contribute
independently to the noise spectrum. This is given by the Fourier transform of
the correlator of $b(t)$,
	\begin{align}
	   S(\nu) = \int d(t-t')\ \llangle b^*(t)b(t')\rrangle e^{i\nu (t-t')}
	\end{align}
At time scales $\tau_\rs{large}$ the correlator is expected to decay because of
the switching to another lasing state. The switching changes the optical phase
but
retains the stationary photon number $n_s=|b_s|^2$. The correlator decays as
	\begin{align*}
	   \llangle b^*(t)b(t')\rrangle = n_s e^{-|t-t'|/\tau_\rs{large}}
	\end{align*}
The  noise spectrum of this correlator is a Lorentzian with exponentially small
width, $1/\tau_\rs{large}$ and with peak height $S(0)= 2n_s \tau_\rs{large}$. At
time scales $\tau_\rs{small}$ small fluctuations have a dominating contribution
to the correlator. Writing $b(t)=b_s + a(t)$ for a small fluctuation, $a(t)\ll
b_s$, we have $\llangle b^*(t)b(t')\rrangle = \langle a^*(t)a(t')\rangle$. The
requirement that the fluctuations $a(t)$ are small implies that they are
Gaussian distributed as shown below. One expects the noise
spectrum to consist of three peaks. One peak is associated with the fluctuations
in detuning, and has the width $\simeq\gamma$. Two side peaks, shifted with
respect to zero detuning, are associated with the intrinsic quantum
fluctuations and have width $\simeq 1$. These peaks are
shifted because of the oscillatory behavior of small deviations from
equilibrium position, which was described in the previous section in the
context of stability analysis. In fact, the shifts should be
$\pm\sqrt{4(A^2-1)(1+\chi)-1}$. The noise spectrum approximately
corresponds to a sum of three Lorentzians. Here the peak heights must be
$S(0)\simeq 2\langle|a(0)|^2\rangle \tau_\rs{small} \ll 2n_s \tau_\rs{large}$.
Adding up the two contributions, the
total noise spectrum is a high and very narrow peak on a background formed by
three low and broad peaks, as depicted in Fig.~\ref{fig:noise}.
In this section we concentrate on the small fluctuations that are
responsible for the broad background in the noise spectrum of the figure.

A full description of the small fluctuations can be found by linearizing the
Fokker-Planck equation of Eq.~\eqref{eq:FPE}. This amounts to expanding $b$ to
$b(t)=b_s + a(t)$ and only retaining up to linear terms in $a$, $a^*$, and
$\tilde\chi$. The resulting Fokker-Planck equation can be solved
exactly,\cite{vankampen} but we do not do this, because we are only
interested
in the steady state probability distribution. The latter is fully characterized
by the variances of the variables of the system.

Instead of directly using the linearized Fokker-Planck equation of
Eq.~\eqref{eq:FPE} to calculate the variances, we use the equivalent system of
Langevin equations for a set of appropriate variables. The advantage is that in
the course of this procedure, we directly obtain the noise spectrum associated
with the linear noise. The linearized form of
Eq.~\eqref{eq:FPE} is fully equivalent\cite{vankampen} to the following system
of Langevin equations, 
	\begin{align}
		\dot a = &-\left[ i(\sqrt{A^2-1}\chi +2\Omega''n_s) + 1 \right]
a\nonumber\\
				&-i \left( A + \Omega''b_s^2 \right) a^*-i\tilde\chi(t)
				\sqrt{A^2-1}\ b_s + \xi_b(t), \\
		\dot{\tilde\chi} = &- \gamma\tilde\chi + \xi_{\tilde\chi}(t).
			\nonumber
	\end{align}
Here it is noted that the first equation is complex and represents two
equations. There are two different Langevin forces satisfying
	\begin{align}\label{eq:noisecorrelations}
		\langle  \xi_b(t)\rangle &= 0,& 
		\langle  \xi_b^*(t) \xi_b(t')\rangle &= \delta(t-t'), \\
	   	\langle\xi_{\tilde\chi}(t)\rangle &= 0 & 
		\langle \xi_{\tilde\chi}(t)\xi_{\tilde\chi}(t')\rangle &= 
			2\gamma\langle \tilde\chi^2\rangle \delta(t-t'), \nonumber
	\end{align}
with other correlators being zero. The source $\xi_b(t)$ represents the quantum
noise in the resonator mode, while the source
$\langle\xi_{\tilde\chi}(t)\rangle$
represents the noise in the detuning. The current set of equations is not the
most convenient one because the variables $a$ and $a^*$ are not orthogonal to
each other. It will be most natural to work in a system with a variable $\delta
n$, representing radial displacements or photon number changes, and a variable
$\delta\varphi$, representing azimuthal displacements or phase changes. The
relation between these two bases comes about in the following way.
One can write the deviation $a = \delta x + i\delta y$ in vector form in a
Cartesian coordinate system as $\vec{a}=\delta x\ \hat x +\delta y\ \hat y$.
Alternatively, in a cylindrical coordinate system $\vec{a} = \delta r\ \hat r +
\sqrt{n_s}\delta\varphi\ \hat\varphi$, where the $\sqrt{n_s}$ follows because
$a$ is taken about $b_s$, which is separated from the origin with a distance
$\sqrt{n_s}$. The variations $\delta\varphi$ represent the variations in
optical phase, but the variations $\delta r$ correspond to the fluctuations of
the square root of $n$. Hence, $\delta r =\delta n/(2\sqrt{n_s})$. Using the
transformation relations, $\hat r = \cos\varphi\ \hat x
+ \sin\varphi\ \hat y$ and $\hat \varphi = -\sin\varphi\ \hat x + \cos\varphi\
\hat y$, one can derive expressions for $\delta x$ and $\delta y$ to find that
	\begin{align*}
	   a = \left[\frac{\delta n}{2\sqrt{n_s}}+ i\sqrt{n_s}
				\delta\varphi\right]e^{i\varphi}
	\end{align*}
Using this we can rewrite the above set of Langevin equations. Reminding
the reader that we use for the lasing solution $n_s=n_-$ and
$\varphi_s=\varphi_-$ [Eq.~\eqref{eq:scsoln}], we arrive at
	\begin{align}\label{eq:langevineqs}
	   \dot{\delta n} &= -4n_s\sqrt{(A^2-1)}\ \delta\varphi + \xi_n(t),
		\nonumber\\
		\dot{\delta\varphi} &= -2\delta\varphi +\frac{1}{n_s}
			\sqrt{A^2-1}(1+\chi)\delta n   \\
			&\qquad -\sqrt{(A^2-1)}\tilde\chi + \xi_{\varphi}(t), \nonumber\\
		\dot{\tilde\chi} &= - \gamma\tilde\chi + \xi_{\tilde\chi}(t).
			\nonumber
	\end{align}
Here we have three Langevin forces satisfying
	\begin{align}
		\langle  \xi_n(t)\rangle &= 0,& 
		\langle  \xi_n(t) \xi_n(t')\rangle &= n_s\delta(t-t'),
		\nonumber \\
	   	\langle\xi_{\varphi}(t)\rangle &= 0 & 
		\langle \xi_\varphi(t)\xi_\varphi(t')\rangle &= 
			\frac{1}{4n_s} \delta(t-t'), \\
	\langle\xi_{\tilde\chi}(t)\rangle &= 0 & 
		\langle \xi_{\tilde\chi}(t)\xi_{\tilde\chi}(t')\rangle &= 
			2\gamma\langle \tilde\chi^2\rangle \delta(t-t'). \nonumber
	\end{align}
The other correlators are zero.

With the change of variables, we now need to find the steady-state probability
distribution to the Fokker-Planck equation for the variables $\delta n$,
$\delta\varphi$, and $\tilde\chi$. This is given by\cite{vankampen}
	\begin{align}\label{eq:probdistr}
	   P_s(\delta r,\delta\varphi,\tilde\chi) &=
		\frac{1}{\sqrt{(2\pi)^3\text{Det}(\Xi)}}
		e^{-\frac{1}{2}\vec{v}^T\cdot \Xi^{-1}\cdot\vec{v}}, \\
		\vec{v}^T &= (\delta n, \delta\varphi,\tilde\chi), \nonumber \\
		\Xi &= 
			\begin{pmatrix} 
				\langle\delta n^2\rangle & \langle\delta n\delta\varphi\rangle  &
									\langle \delta n\tilde\chi\rangle\\
				\langle\delta n\delta\varphi\rangle &\langle\delta\varphi^2\rangle&
									\langle \delta \varphi\tilde\chi\rangle\\
				\langle \delta n\tilde\chi\rangle & \langle \delta
					\varphi\tilde\chi\rangle  & \langle\tilde\chi^2\rangle\\
			\end{pmatrix},\nonumber
	\end{align}
where $\langle\tilde\chi\rangle$ is simply the variance of the distribution
of the detuning fluctuations. Now using the Langevin equations we can find the
variance matrix $\Xi$ according to the following procedure. First, we take the
Fourier transform of Eqs.~\eqref{eq:langevineqs}. For any
variable $g(t)$ we have $g(t)=\int d\nu\ g_\nu \exp[-i\nu]$. The resulting
system of equations is solved for $\delta n_\nu$, $\delta\varphi_\nu$ and
$\tilde\chi_\nu$. To get the variances, we calculate the inverse
Fourier transform at $t=0$ of the expectation value of products of the
Fourier transformed variables. For instance, we have
	\begin{align}
	   \langle \delta n \delta\varphi \rangle =	\frac{1}{2\pi}
			\int_{-\infty}^\infty d\nu\langle\delta n_\nu\delta\varphi_\nu \rangle.
	\end{align}
According to this procedure, we find the variances and write them in a
form that will be convenient later:
	\begin{align}\label{eq:variances}
	   \frac{\langle\delta n^2\rangle}{n_s^2} &= \frac{2(\gamma+2)(A^2-1)
			\langle\tilde\chi^2\rangle} {(1+\chi)[4(A^2-1)(1+\chi)+ \gamma^2 +
			2\gamma] } \nonumber \\
			&+\frac{1}{4n_s} \left[1+ \frac{A^2}{(A^2-1)(1+\chi)}\right],
			\nonumber\\
		\langle\delta\varphi^2\rangle &= 			
			\frac{\gamma(A^2-1)\langle\tilde\chi^2\rangle}					
			{2[4(A^2-1)(1+\chi)+ \gamma^2 +2\gamma] }
			+ \frac{2+\chi}{16n_s}, \nonumber \\
		\frac{\langle\delta n\delta\varphi\rangle}{n_s} &=
					\frac{1}{4n_s\sqrt{(A^2-1)}}, \\
		\frac{\langle \delta n\tilde\chi\rangle}{n_s} &= 
			\frac{4(A^2-1)\langle\tilde\chi^2\rangle}
				{4(A^2-1)(1+\chi)+ \gamma^2 + 2\gamma },\nonumber \\
		\langle \delta \varphi\tilde\chi\rangle &= 
			\frac{-\gamma \sqrt{A^2-1}\langle\tilde\chi^2\rangle} 
				{4(A^2-1)(1+\chi)+ \gamma^2 + 2\gamma }.\nonumber
	\end{align}
We remark that some variances diverge when approaching the lasing threshold,
which is when $|A|\to1$ or $\chi\to-1$. Also, the combination
$4(A^2-1)(1+\chi)$ is equal to the determinant of the stability matrix of
Eq.~\eqref{eq:stability}. A larger value of the determinant implies higher
stability, as the variances decrease.

With the expressions for the variances, we are able to quantify in which
parameter regimes the linear Fokker-Planck equation and the steady-state
distribution of Eq.~\eqref{eq:probdistr} are valid. Their validity is ensured
when $\delta n\ll n_s$ and $\delta\varphi\ll1$, so that all expressions in
Eq.~\eqref{eq:variances} must be much smaller than one. Note that under these
conditions, there is no {\it a priori} bound on $\tilde\chi$, which is already
purely
linear. It will be particularly useful to study the expression for
$\langle\delta n^2\rangle/n_s^2$ in a bit more
detail, because it is always of the order of, or larger than the other ones.
Additionally, this is an interesting quantity, because large fluctuations in
$\delta n$ lead to switchings and are thus important for the long coherence
times.

The variance of the fluctuations in photon number is studied in a few limits.
For reasons of simplicity and clarity, it is most convenient to be
sufficiently far from the
lasing threshold, so we take $\chi\ll1$ and $A\gg1$. Keeping this in mind, the
first limit to be studied is the one where $A\gg\gamma$. Then we have
	\begin{align}\label{eq:Agggamma}
	   \frac{\langle\delta n^2\rangle}{n_s^2} = \frac{1}{2}(\gamma+2)
			\langle\tilde\chi^2\rangle + \frac{1}{n_s}.
	\end{align}
We note a few properties. To start with, the term proportional to
$\langle\tilde\chi^2\rangle$ is associated with the noise in the detuning,
whereas the $1/n_s$ term is associated with the intrinsic quantum noise of the
resonator mode. The latter exactly corresponds to the fundamental
noise expected in any common laser.\cite{scully} In line with this division,
there are two
distinct regimes when considering the dependence on $n_s$. For small enough
values of $n_s$ the intrinsic quantum noise dominates so that the variance
decreases inversely proportional to $n_s$. Increasing $n_s$ beyond a critical
value the variance saturates as the $1/n_s$ term becomes unimportant and the
noise in detuning starts to dominate. We further note that the magnitude of the
fluctuations is independent of $A$ and independent of $\gamma$ in the limit
$\gamma\ll1$. The criterion to have small fluctuations now leads to the
requirements that $n_s\gg1$ and $(\gamma+2)\langle\tilde\chi^2\rangle\ll1$.

The two previously mentioned limits of $\gamma$ in the detuning noise dominated
regime can be intuitively understood. If $\gamma\ll1$, it is smaller than
the decay rate of the resonator mode so that the fluctuations in $\tilde\chi$
are slow enough for the optical field to adiabatically follow them. The number
fluctuations are then purely determined by the equilibrium dipole distribution.
When $\gamma\gg1$, the optical field cannot follow the dipole fluctuations
adiabatically. However, because $A\gg\gamma$, the optical field will respond
fast to the dipole fluctuation before it fades away. In fact, the optical
field will start to rotate about the stationary point with a period
$t\sim2\pi/A$ and the fluctuation only fades away after $\sim A/2\pi$ rotations
about the stationary lasing state.
The variance of the number fluctuations is now determined by the instantaneous,
out-of-equilibrium value of the fluctuation of $\tilde\chi$, as determined by
$\xi_{\tilde\chi}(t)$ in Eq.~\eqref{eq:noisecorrelations}, leading to an extra
factor of $\gamma$ compared to the previous case with $\gamma\ll1$.

We continue by studying the variance of photon number in the second limit, which
is $A\ll\gamma$ and leads to
	\begin{align}\label{eq:Allgamma}
	   \frac{\langle\delta n^2\rangle}{n_s^2} = \frac{2A^2}{\gamma}
			\langle\tilde\chi^2\rangle + \frac{1}{n_s}.
	\end{align}
In this limit, again there is in the dependence on $n_s$ a division in two
regimes: one where the variance is dominated by the noise in detuning and one
where the intrinsic quantum noise of the resonator mode is most important. In
contrast to previous case there is now a dependence on both $A$ and $\gamma$.
Here, the criterion to have small fluctuations leads to the requirements that
$n_s\gg1$ and $A^2\langle\tilde\chi^2\rangle\ll\gamma$.

There is also an intuitive way to understand the detuning noise-dominated regime
in the limit $A\ll\gamma$. Here, the dynamics of the detuning fluctuations
is on much shorter time scales, $1/\gamma$, than the dynamics of the resonator
mode, $1/A$, so that the latter has a reduced time window to respond to the
detuning fluctuations. Compared to the earlier case with $1\ll\gamma\ll A$, the
reduction factor is $4A^2/\gamma^2$.

As part of the calculation done to find the variances, we can also find an
expression for the noise spectrum, which is defined as $S(\nu)=\langle|\delta
a_\nu|^2\rangle = \langle|\delta n_\nu|^2\rangle/(4n_s) + n_s\langle|
\delta\varphi_\nu |^2\rangle$. Defining $D = 4(A^2-1)(1+\chi)$ we have
\begin{widetext}
	\begin{align}\label{eq:noisespectrum}
	   S(\nu) = \frac{1}{4[(\nu+\sqrt{D-1})^2+1][(\nu-\sqrt{D-1})^2+1]}
			\left[4A^2+2\nu^2 + D(1+\chi)+
				\frac{8n_s(A^2-1)[4(A^2-1)+\nu^2]\gamma \langle\tilde
			\chi^2\rangle} {\gamma^2+\nu^2} \right] 
	\end{align}
\end{widetext}
The term proportional to $\langle\tilde\chi^2\rangle$ is directly
related to the noise in the detuning, whereas the other term is directly related
to the quantum noise of the resonator mode. Which of the two terms is the most
important depends, of course, on the parameter regime, as discussed above. The
noise spectrum can be written in three parts, as we show in the Appendix and
discussed in the intuitive picture above, with each part corresponding to a
peak. There is a central Lorentzian peak at $\nu=0$ with width $2\gamma$ and
there are two non-Lorentzian sideband peaks at $\nu_\pm=\pm\sqrt{D-1}$. In the
limit $A\gg1$ and $\chi\ll1$ the sideband peaks become Lorentzian with width
$2$. In this limit, the central peak typically dominates over the sideband
peaks when $\gamma\ll A$. For $\gamma\gg A$, the central peak almost vanishes
compared to the sideband peaks.

We have now described the linear fluctuations in the HJL and derived conditions
on parameters for which fluctuations are small, $\langle\delta
n^2\rangle/n_s^2\ll1$. The noise in various parameter regimes was studied. For
small $n_s$, but still much larger than $1$, the intrinsic quantum noise of the
resonator mode was found to have a dominating contribution to the photon number
fluctuations, while for large $n_s$, the noise in the detuning is the most
important. There, the magnitude of number fluctuations depends on $\gamma$ and
the ratio $A/\gamma$. The noise spectrum corresponding to the small
fluctuations was found and has a central peak with varying width, and two
sideband peaks with width $2$.

\section{Large Fluctuations}\label{sec:IV}
In the previous section we studied small fluctuations in the HJL under the
assumption that they dominate the fluctuation spectrum. These small
fluctuations do not cause decoherence of the laser. Decoherence is caused by
large fluctuations that switch the stationary lasing state to one with a phase
difference of $\pi$; see fig~\ref{fig:lasing}. In this section we
study the large fluctuations thoroughly to find the time scales at which
the
switchings would occur. This gives the decoherence time of the HJL.

The probability to have a large fluctuation can be calculated using the
formalism of optimal paths.\cite{reviewop} This formalism is, for instance, used
in classical diffusion driven systems where the transition rate from one
stationary state to another, across a high barrier, is calculated.
The path that crosses the barrier and has the largest probability is called the
optimal path. Its shape and probability can be found using the principle of
least action. Its probability determines the transition rate.

In this section, we introduce the formalism of optimal paths and apply it
to
our system, after which some generic properties of the action are derived. After
this we show how the coherence time is related to the optimal paths. We
investigate the dependence of the action of the optimal paths on the device
parameters. This is done both by using the results of previous section and
by simulating the optimal paths.

\subsection{Trajectories and relation to Kramers' escape problem}
To start with, we outline the dynamics of the device in the absence of noise,
describe the effect of noise and explain the relation to Kramers' escape
problem.

\begin{figure}
  \centering
  \includegraphics{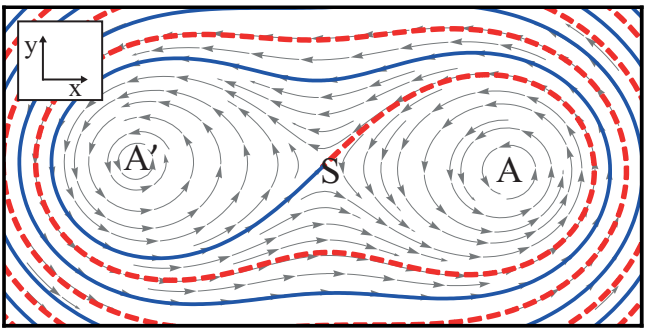}
  \caption{%
(Color online) A stream plot corresponding to Eq.~\eqref{eq:sceqngen} for
variables $x$ (horizontal) and $y$ (vertical), defined according to $b = x+iy$.
The gray streamlines represent a ``force field''. When a ``particle'' is placed
in the force field it will evolve along these lines to one of the attractors,
indicated
by $A$ and $A'$, depending on the initial condition. When placed at the
thick red (dashed) or blue (solid) line, the particle will
not evolve to one of the attractors but to the saddle point, indicated by $S$.
These lines form the separatrix of the system, that separates the domains of
attraction of the attractors.
  }\label{fig:stream}
\end{figure}

In Sec.~\ref{sec:II} we have derived the requirements on stable stationary
lasing in the HJL. Let us suppose that the optical field in the HJL is not
in a point of stable lasing and that $\tilde\chi=0$. If we disregard the noises
we know that the field
will evolve to one of the points. This evolution is described by
Eq.~\eqref{eq:sceqngen}. We visualize the dynamics by making a stream plot, as
is done in Fig.~\ref{fig:stream}. To describe the dynamics it is convenient to
associate the field components, $x,y$, with coordinates of a ``particle''
subject to a coordinate-dependent ``force field'' that causes the motion of the
particle with the velocity proportional to the force. Starting from a given
initial condition, the particle will flow towards a stable stationary point.
Its trajectory corresponds to a streamline in the plot.

Figure~\ref{fig:stream} corresponds to regime II as defined
in Sec.~\ref{sec:II}. In the figure the three stationary points are indicated
by $S$, $A$, and $A'$. The point $S$, at $x,y=0$ is the unstable saddle point,
whereas the points $A$ and $A'$ are stable points or attractors. For a general
initial condition in the vicinity of the saddle point the particle is
repelled from this vicinity. There is, however, a particular line called stable
direction such that the particle is attracted to the saddle point if the
initial condition is chosen to be on this line.  The streamlines coming to the
saddle point at the stable direction form the separatrix of
Eq.~\eqref{eq:sceqngen}, which is the boundary separating the domains of
attraction of the attractors $A$ and $A'$. We see that the particle circles
around an attractor before reaching it. This signifies that the force field
cannot be represented as a gradient of a scalar
potential. The force field must also have a transversal component
that causes the rotation and can be represented as the curl of a vector
potential. 

Full equations of motion shall also include a variation of detuning
$\tilde\chi$. With this we have three coupled equations of motion
that define the force field in the space of three variables,
	\begin{align}\label{eq:3deqsmotion}
		\begin{split}
			\dot x &= -x \left[\sqrt{A^2-1}(\chi+\tilde\chi) -A +  
					\Omega'' n\right]y, \\ 
			\dot y &= -y -\left[\sqrt{A^2-1}(\chi+\tilde\chi) +A +  
					\Omega'' n\right]x \\
			\dot{\tilde\chi} &= -\gamma\tilde\chi.
		\end{split}
	\end{align}
The last equation gives the relaxation of the detuning to its equilibrium
position. The specifics of this equation is the presence of a {\it saddle
line}. 
The line $x,y=0$ is a saddle line because for each value of $\chi$ the point
$x,y=0$ is a saddle point with respect to the motion in $x,y$ directions.

In Sec.~\ref{sec:III} noise was introduced. With the Langevin noise added to
the dynamics, Fig.~\ref{fig:stream} no longer gives a full
description of the particle evolution. The particle in the `force
field' will experience random kicks, causing it to change from one trajectory to
another, even when initially at rest at an attractor. Effectively, the kicks
cause the particle to diffuse away from its starting point, even against the
streamlines of the force field. Nonetheless, the force field counters the
diffusive flow trying to bring
the particle back to the attractor. These two kinds of motion have a
fundamentally different nature. The evolution along the flow lines is
deterministic, in contrast to the diffusive motion which is nondeterministic.
Importantly, the diffusive motion of the particle lifts up the
restriction of the particle to stay in the domain of attraction of an
attractor.
A series of random kicks may repel the particle from an attractor and
eventually pull it over the separatrix into the domain of
attraction of the other attractor. After that the particle relaxes naturally
along a streamline to the other attractor. This trajectory corresponds to the
large fluctuation we are after: A switch between the two stable lasing states
has occurred.

Let us suppose, in contrast to what is the case for Eq.~\eqref{eq:sceqngen},
that a particle moves in a potential well, of which the gradient defines the
corresponding force field, and that the fluctuation-dissipation theorem is
applicable, like in Kramers' escape problem. In this case the Fokker-Planck
equation can be solved to calculate the escape probability from the potential
well.\cite{kramers,vankampen} This escape probability involves the Arrhenius
factor, $\exp(W/T)$, where $W$ is the height of the barrier that is crossed by
the particle, counted from the potential minimum, and $T$ is the effective
temperature of the particle. However, because of the transversal component of
the force field associated with Eq.~\eqref{eq:sceqngen}, we cannot use this
general approach to solve Fokker-Planck equation~\eqref{eq:FPE}. There, the
escape probability does not depend on a potential barrier height, but on the
shape and probability of the path that crosses the separatrix. We note that
the potential barrier and the separatrix both form the boundary of the domain
of attraction where the particle is trapped.

\subsection{Optimal paths and the principle of least action}
\label{sec:actionHJL}
With the conceptual framework in mind, let us continue by showing how the
probability to have a transition from one stationary state to the other, is
related to the concept of the optimal path. We only consider paths that
start in stationary point $A$ and end in stationary point $A'$. The time
elapsing during the transition will be infinite, going from $-\infty$ to
$\infty$, because the particle can only leave $A$ and approach $A'$
asymptotically.

It is possible to express the probability distribution for the fluctuations in
terms of an action. We start with a general set of Langevin equations, written
in vector form,
	\begin{align}\label{eq:langevin}
	   \dot{\vec{r}} = \vec{V}(\vec{r}) + \vec{\xi}(t).
	\end{align}
The Langevin forces $\xi_i(t)$ represent Gaussian white noise, satisfying
$\langle\xi_i(t)\rangle=0$ and $\langle\xi_i(t)\xi_j(t')\rangle =
D_i\delta_{ij} \delta(t-t')$, thus supposing an infinitesimally short
autocorrelation time. The noise can be expressed as a single
probability distribution that is a product of the probability
distributions of the separate Langevin forces at each moment in time,
	\begin{align}\label{eq:noiseprobability}
	   P = C \exp\left[ -\int \sum_i \frac{\xi_i(t')^2}{2D_i}
			dt' \right]
	\end{align}
with $C$ the normalization constant. One may interpret this expression in
another perspective, as the probability distribution for histories of
fluctuations, because the integral in the exponent represents the history of
fluctuations. Each history corresponds to a unique progress in time of
$\vec{\xi}(t')$. We stress that the values
of $\vec{\xi}(t)$ at each time $t$ remain strictly independent from the ones at
other times $\tilde t\ne t$. So how does this concept of histories aid us? Even
though
the Langevin forces might have infinitesimally short autocorrelation times, the
vector field $\vec{r}$ will need some finite time to relax to the stationary
state. During this time, the history of fluctuations will be relevant. The
resulting cumulative effect of fluctuations is best captured by using
Eq.~\eqref{eq:langevin} to rewrite the exponent of $P$, yielding
	\begin{align}\label{eq:definitionaction}
	   P = C e^{-S}, \quad S = \int 
			\sum_i \frac{(\dot r_i-V_i)^2}{2D_i} dt',
	\end{align}
Hence, we indeed see that the noise probability distribution can
be described using the action, defined as $S$.

The probability distribution, $P$, is strongly reminiscent of a path
integral. Indeed, $P$ is a probability density which, upon integration over all
paths, gives the probability to go from $A$ to $A'$. When fluctuations are
small, as is described in Sec.~\ref{sec:III}, the path integral is dominated by
the ``classical'' path and by the quadratic fluctuations about
it.\cite{kleinert}
The classical path is found by the principle of least action. It is what we
have called the optimal path. The optimal path corresponds to the lowest
value of the action for the transition, $S_\text{opt}$, which relates to the
transition probability
	\begin{align*}
		P \simeq F e^{-S_\text{opt}} 
	\end{align*}
with $F$ the prefactor corresponding to the quadratic fluctuations. Hence, the
probability of a transition from $A$ to $A'$ depends on the optimal path.

Without specifying any details about $\vec{V}$ we note two things. First, by
inspection we see that $S\geq0$, with the equality satisfied for the
deterministic path, for which $\dot{\vec{r}} = \vec{V}$. Second, the integrand
of $S$ is equivalent to a Lagrangian and because it only implicitly depends on
time, a conserved quantity similar to the Hamiltonian exists
	\begin{align}\label{eq:consham}
	   H = \sum_i\frac{1}{2}\left[ \left(\frac{\dot{r}_i}{D_i}\right)^2 
			- \left(\frac{V_i}{D_i}\right)^2\right]=0.
	\end{align}
The equality of the Hamiltonian to zero is not universal, but is appropriate for
the path we consider. This path connects the stationary points, where
$\dot{\vec{r}}=\vec{V}=0$ and therefore $H=0$. Being a conserved quantity, $H$
must always be zero at the path connecting the stationary points.

The action as it is given in Eq.~\eqref{eq:definitionaction} is not in the
most convenient form. To minimize this action we need to find both the shape of
the optimal path and the speed with which the path is traveled. By exploiting
the conserved Hamiltonian, we can relate the speed to the force field
resulting in an action where the optimal path is only determined by its shape.
To show this, we choose the diffusion constants, without losing generality, as
$D_i=1$ for all $i$. This is allowed, because we can always rescale $r_i$ and
$V_i$ accordingly. Applying this to Eq.~\eqref{eq:consham} we then find that
$|\dot{\vec{r}}| = |\vec{V}|$. With this, the terms $\dot{\vec{r}}^2/2$ and
$\vec{V}^2/2$ can be rewritten and added to each other to become
$|\dot{\vec{r}}| |\vec{V}|$. Because both terms in the Lagrangian now only
contain a single time derivative, we can parametrize time at will. We
parametrize it as a monotonously growing function $\zeta'(t)>0$, such that
$\zeta(t_0)=0$ and $\zeta(t_f)=1$. Consequently $\dot r_i = \dot\zeta (\partial
\vec{r}/\partial\zeta)$, so that the action becomes
	\begin{align}\label{eq:transformedaction}
	   S = \int_0^1 \left\{ \left| \frac{\partial \vec{r}}{\partial\zeta}\right|
		 |\vec{V}| - \frac{\partial \vec{r}}{\partial\zeta} \cdot\vec{V} 
			\right\}d\zeta
	\end{align}
Minimizing this action will result in the optimal path irrespective of
the exact form of parametrization $\zeta(t)$ and therefore irrespective of the
speed with which the optimal path is traveled.

From the expressions for the action, Eq.~\eqref{eq:transformedaction}, an
intuitive understanding of the dynamics of the system can be obtained. Suppose
that we are at some point in the force field $\vec{r}$ and move away from this
point a random infinitesimal
distance $d\vec{r}$. The probability to move in a certain direction is then
determined by the increase in action associated with this direction, which is,
in turn, determined by the angle between the force field vectors,
$\vec{V}$ and $d\vec{r}$. With the two vectors aligned the action is smallest
(zero) and the probability to move in that direction the highest. Again, this
corresponds to the deterministic solution. With the vectors counter
aligned the action is largest and the probability to
move in that direction the lowest. Close to stable or unstable stationary
points, when $|\vec{r}|^2$ is of the order of the diffusion constants, the
absolute difference between alignment and counteralignment is small because
$|V|$ is close
to zero. The probabilities to go in any direction become of the same order of
magnitude. This is the regime where the quadratic fluctuations are important.
Far away from stationary points, however, $|V|$ can be quite large so that the
absolute difference between alignment and counteralignment becomes large. Then,
because the action appears in the exponent, the probability to follow the
deterministic path will be very close to unity. Going in some other direction
will cause a large increase in action.

Concluding this section we make two remarks. First, the
probability distribution of Eq.~\eqref{eq:definitionaction} is a solution to
the Fokker-Planck equation~\eqref{eq:FPE} in the {\it Wentzel-Kramers-Brillouin}
approximation.\cite{kleinert, merzbacher} The solution is exact in the limit
where the diffusion constants are infinitesimally small. Second, the
coherence time, or escape time, is related to the transition
probability,\cite{gardiner} yielding
	\begin{align}
	   \tau_\rs{esc} = K e^{S_\text{opt}}
	\end{align}
Here, $K$ relates to the quadratic fluctuations around the optimal path. We
disregard this prefactor and only consider the action of the optimal path,
which determines the order of magnitude of the coherence time.

\subsection{The action for the HJL}
Let us apply the above described ideas and calculate the action for the HJL,
using the Fokker-Planck equation of Eq.~\eqref{eq:FPE}. The Fokker-Planck
equation contains three variables. Two of them, $x,y$, associated with the
optical
field, defined as $b=x+iy$, and one associated with the fluctuations in
detuning,
$\tilde\chi$. We redefine the latter as $z\equiv\tilde\chi /
\sqrt{D_{\tilde\chi}}$, where $D_{\tilde\chi}= 2\gamma
\llangle\tilde\chi^2\rrangle$ is the diffusion constant of
$\tilde\chi$. As a result, the diffusion constants associated with fluctuations
in
$x,y,z$ are all equal to one. Because any Fokker-Planck equation, with constant
diffusion terms, has an equivalent Langevin equation,\cite{vankampen}
Eq.~\eqref{eq:FPE} can be written in the form of Eq.~\eqref{eq:langevin}, with
	\begin{align}\label{eq:langevinspecific}
		\begin{split}
	   \vec{V} &= 
			\begin{pmatrix}
			   -x+\chi\sqrt{A^2-1}\ y+\frac{1}{2}\frac{\partial U}{\partial y} \\
				-y-\chi\sqrt{A^2-1}\ x-\frac{1}{2}\frac{\partial U}{\partial x} \\
				-\gamma  z
			\end{pmatrix}\\
			U &= \Omega + \sqrt{2\gamma\llangle\tilde\chi^2\rrangle
			(A^2-1)}\, z(x^2+y^2) \\
			&\qquad\qquad +\frac{1}{2}\Omega''(x^2+y^2)^2 + A(x^2-y^2) \\
			D_x & =D_y=D_z=1
		\end{split}
	\end{align}
where $\vec{r}=(x,y,z)$ is a dimensionless quantity. We remind the reader that
$A$,
$\gamma$, $\omega$, $\Omega$, $\Omega'$, $\Omega''$ and $U$
[Eq.~\eqref{eq:energy}] are dimensionless with the corresponding
frequencies being measured in units of $\Gamma/2$. The drift terms $\vec{V}$ can
be explicitly written in terms of scalar and vector potentials
	\begin{align}
		\begin{split}
	   \vec{V} &= -\nabla\phi + \nabla\times\vec{A}, \qquad 
		\phi = \frac{1}{2}(x^2 + y^2) +\frac{\gamma}{2}z^2, \\
		\vec{A}&= A_z\hat z, \qquad A_z = \frac{1}{2}[\omega(x^2+y^2)+U],
		\end{split}
	\end{align}
demonstrating that the vector field corresponding to
$\vec{V}$ contains both longitudinal and transversal parts. The action for the
HJL is now simply given by Eq.~\eqref{eq:transformedaction} with the vector and
the diffusion constants given by Eq.~\eqref{eq:langevinspecific}.

\subsection{Estimation of the action}
Now that we have discussed the formalism of optimal paths and have established
the relation with the escape time, we want to find explicitly the value of the
optimal action, depending on the system parameters of the HJL. Before that, let
us present the estimation of the action for the optimal path, $S_\text{opt}$,
in different regimes. For this we use the result of Sec.~\ref{sec:III}. In
the following section we simulate numerically the optimal paths and the
corresponding action, to validate the findings of this section.

In Sec.~\ref{sec:III} we have discussed the fluctuations in the device in
the linear noise approximation. Under these assumptions the fluctuations are
small and obey a Gaussian distribution with variances given by
Eq.~\eqref{eq:probdistr}.

For the order-of-value estimations we can use this formula for small
fluctuations to find the probability of large fluctuations. In particular, large
fluctuations in $n$ are important, because at $n=0$, the particle reaches the
saddle point enabling the switching to the opposite stable point. Large
fluctuations in either of the other variables will not drive the
particle across the separatrix. The probability for this fluctuation in
Gaussian approximation is simply proportional to $\exp[-n_s^2/(2\langle\delta
n^2\rangle)]$, with $\langle\delta n^2\rangle$ given by
Eq.~\eqref{eq:probdistr}
	\begin{align}\label{eq:optimalaction}
	   \frac{\langle\delta n^2\rangle}{n_s} &= \frac{2(\gamma+2)(A^2-1)
			\langle\tilde\chi^2\rangle} {(1+\chi)[4(A^2-1)(1+\chi)+ \gamma^2 +
			2\gamma] } \nonumber \\
			 &+\frac{1}{2n_s} \left(1+ \frac{A^2}{(A^2-1)(1+\chi)}\right)
	\end{align}
We can thus estimate $S_\text{opt}\simeq n_s^2/\langle\delta
n^2\rangle$, where the coefficient is different in different parameter regimes.
Let us list here all possible limiting cases corresponding to different
relations between parameters of the device. 

In the case (i) the large fluctuation is caused by quantum photon noise. The
fluctuation is estimated by the last term in the
expression~\eqref{eq:optimalaction} and the optimal action is given by
	\begin{align}\label{eq:case(i)}
	   S_\text{opt}=n_s(1+\chi) F_1(A,\chi)
	\end{align}
$F_1$ being a coefficient of the order of one. The probability of the large
fluctuation is enhanced near the lasing threshold, $\chi=-1$, and is
exponentially small otherwise, provided $n_s\gg1$. 

In all other cases the
probability of the large fluctuation is mainly caused by the detuning noise. In
the limit $\gamma\ll1$, when the switching time of the emitters is longer than
the damping time of the optical resonator, we can distinguish two cases. The
case (ii) takes place near the lasing threshold, $A\approx 1$, and is
characterized by $\gamma\gg A^2-1$. The estimation for $S_\text{opt}$ reads
	\begin{align}\label{eq:case(ii)}
	   S_\text{opt}= \frac{\gamma(1+\chi)}{(A^2-1)\langle\tilde\chi^2\rangle}
			F_2(\chi)
	\end{align}
and is inversely proportional to the variance of the detuning fluctuations. In
the opposite case (iii) $\gamma\ll A^2-1$. The estimation is given by
	\begin{align}\label{eq:case(iii)}
	   S_\text{opt}= \frac{(1+\chi)^2}{\langle\tilde\chi^2\rangle}
			F_3(A,\chi)
	\end{align}
and does not depend on $\gamma$. 

If the switching times are relatively short, $\gamma\gg 1$, we distinguish two
cases depending on the ratio of $\gamma$ and $A$. In case (iv) $\gamma\gg A$
and the estimation reads
	\begin{align}\label{eq:case(iv)}
	   S_\text{opt}= \frac{\gamma(1+\chi)}{A^2\langle\tilde\chi^2\rangle}
			F_4(\chi)
	\end{align}
while in the opposite case (v) ($\gamma\ll A$) it reads
	\begin{align}\label{eq:case(v)}
	   S_\text{opt}= \frac{(1+\chi)}{\gamma\langle\tilde\chi^2\rangle}
			F_5(\chi).
	\end{align}
In all these cases coefficients $F_{2-5}$ are supposed to be of the order of
one. The boundaries in parameter space separating case (i) from cases (ii)-(v)
are found by comparing the corresponding estimations.

\subsection{Numerical results}
Now that we have the estimations of the optimal action and the understanding of
its dependence on relevant parameters, we quantify numerically the shape
of the optimal path and the action. We describe the numerical method and we
comment briefly on its limitations. We compare the numerical results and the
estimations and discuss the peculiarities of the path shapes.

We use the action of Eq.~\eqref{eq:transformedaction}. A path $\vec r(\zeta)$
starts at the attractor at positive $x$, negative $y$ and $z=0$ ($A$ in
Fig.~\ref{fig:stream}) and ends at the point $A'$. The path can be separated
into two parts. During the first part the path proceeds from $A$ untill a point
at the separatrix that divides the domains of attraction. During the second part
the path follows the natural motion of the particle satisfying equations of
motion Eq.~\eqref{eq:3deqsmotion}. This part does not contribute to the action,
since $d\vec{r}/d\zeta\parallel \vec{V}$, and thus does not have to be evaluated
numerically. In all cases investigated numerically, we have found that the first
part of the path ends at the saddle line $x,y=0$ at some yet to be optimized
value of $z$. 

To find the optimal path we could first formulate a boundary value problem,
using the differential Euler-Lagrange equation from the action
Eq.\eqref{eq:langevinspecific}. The solution to this is the optimal path,
which can only be found using a shooting method. With this method, one changes
initial conditions and integrates the Euler-Lagrange equation trying to reach 
the predefined final condition. This method is not efficient to find
the trajectories we are looking for because they have a relatively low
probability.

We implement a more efficient method. We directly minimize the action of a
predefined path, that depends on a finite number of
parameters, as function of the path parameters. For the path parametrization, we
implement a Bezier curve of some sufficiently high order $n$ (from 12 to 16
depending on the region in the parameter space),
	\begin{align}
	   r(\zeta) = \sum_{i=0}^n \zeta^i(1-\zeta)^{n-i} \vec{P}_i.
	\end{align}
The order plus one equals to the number of points $\vec{P}_i$ in $xyz$ space
that define
the curve. The starting point of the Bezier curve, $\vec{P}_0$, is fixed to the
attractor while the last point, $\vec{P}_n$ is fixed to the saddle line, $P_n^z$
being the subject of the optimization procedure, as well as the coordinates of
all other points. The accuracy of this method
depends on the order of the Bezier curve used. The higher the order, the better
the accuracy. The order required to have a sufficiently accurate path will
depend on the complexity of the shape of the path. 

\begin{figure*}
  \includegraphics{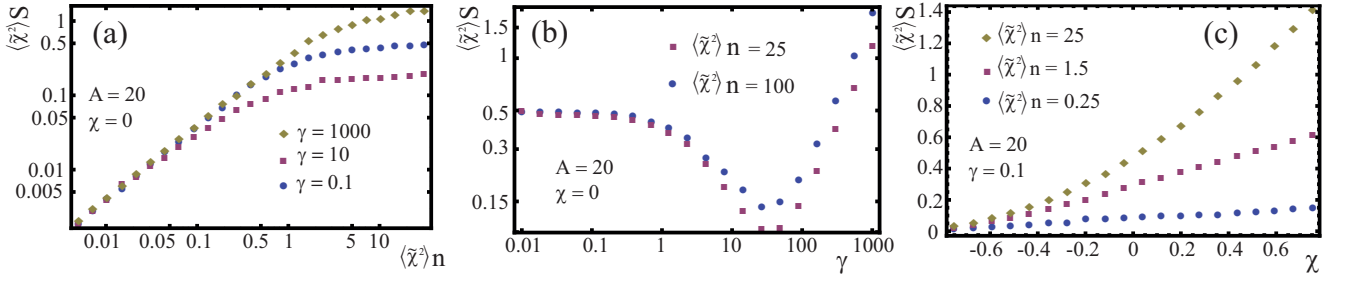}
  \caption{%
  (Color online) The value of the action of the optimal path for several
parameter regimes. Fixed parameters are denoted in the graphs. Parameters were
chosen such that we are far away from the lasing threshold where the use of
optimal paths is justified. The main text contains a comparison with
Eqs.~\eqref{eq:case(i)}-\eqref{eq:case(v)} to estimate the values $F_{1-5}$.
(a) The action as a function of $n \langle\tilde\chi^2\rangle$, for three
values of $\gamma$. The axes are logarithmically
scaled.  For small values of $n \langle\tilde\chi^2\rangle$ the action is linear
in photon number and independent of $\gamma$. This is the regime where intrinsic
photon number fluctuations dominate. For large $n \langle\tilde\chi^2\rangle$
the detuning fluctuations dominate and the action saturates to a value depending
on $\gamma$. 
(b) The action as a function of $\gamma$ for two values of $n
\langle\tilde\chi^2\rangle$, with logarithmically scaled axes. For $\gamma\ll1$
the $\langle\tilde\chi^2\rangle S$ is constant, while for $\gamma\gg A$ it is
proportional to $\gamma$. In the region
between we see a minimum at $\gamma\simeq A$. The curve for $n
\langle\tilde\chi^2\rangle=25$ has points in common with the three curves of
(a). 
(c) The action as a function of the detuning $\chi$, for three values of $n
\langle\tilde\chi^2\rangle$.
  }\label{fig:action}
\end{figure*}

We have calculated the optimal paths for several parameter regimes, to validate
the estimations of Eqs.~\eqref{eq:case(i)}-\eqref{eq:case(v)} and find
numerical values of the prefactors. In Fig.~\ref{fig:action}(a) we show the
dependence of the value of the optimal action on the number of photons
in the resonator mode, $n$, for three different values of $\gamma$ and $\chi=0$.
For sufficiently small values of $n$, $S$ is linear in $n$ and case (i)
applies, yielding $F_1\approx0.4$ for all three values of $\gamma$. For large
values of $n$ the action $S$ saturates to a value depending on $\gamma$. For
$\gamma=0.1$ case (iii) applies, with $F_3\approx0.5$, while case (iv)
applies for
$\gamma=1000$ with $F_4\approx0.5$. The curve for $\gamma=10$ at large $n$
corresponds to a crossover regime between cases (iv) and (v).

In Fig.~\ref{fig:action}(b) we show the dependence of the value of the action on
$\gamma$ for two values of $n \langle\tilde\chi^2\rangle$. In the limit
$\gamma\ll 1$ case (iii) applies, with also here $F_3\approx0.5$. In the limit
$\gamma\gg A$ case (iv) applies with $F_4$ depending on
$\langle\tilde\chi^2\rangle$. At $n \langle\tilde\chi^2\rangle=25$
$F_4\approx0.5$, which is consistent with its value found previously, whereas
at $n \langle\tilde\chi^2\rangle=100$ $F_4\approx0.8$.

Figure~\ref{fig:action}(c) shows the dependence of the value of the action on
the detuning $\chi$ for three values of $n \langle\tilde\chi^2\rangle$. For $n
\langle\tilde\chi^2\rangle=0.25$ case (i) applies with $F_1\approx0.4$, which is
consistent with the value of $F_1$ for Fig.~\ref{fig:action}(a). For $n
\langle\tilde\chi^2\rangle=25$ case (iii) applies with $F_3\approx0.5$,
which is also consistent with the value previously found. The curve with $n
\langle\tilde\chi^2\rangle=1.5$ corresponds to the crossover regime between
cases (i) and (iii) so that no value can be extracted for any of $F_{1-5}$.

\begin{figure*}
  \centering
  \includegraphics{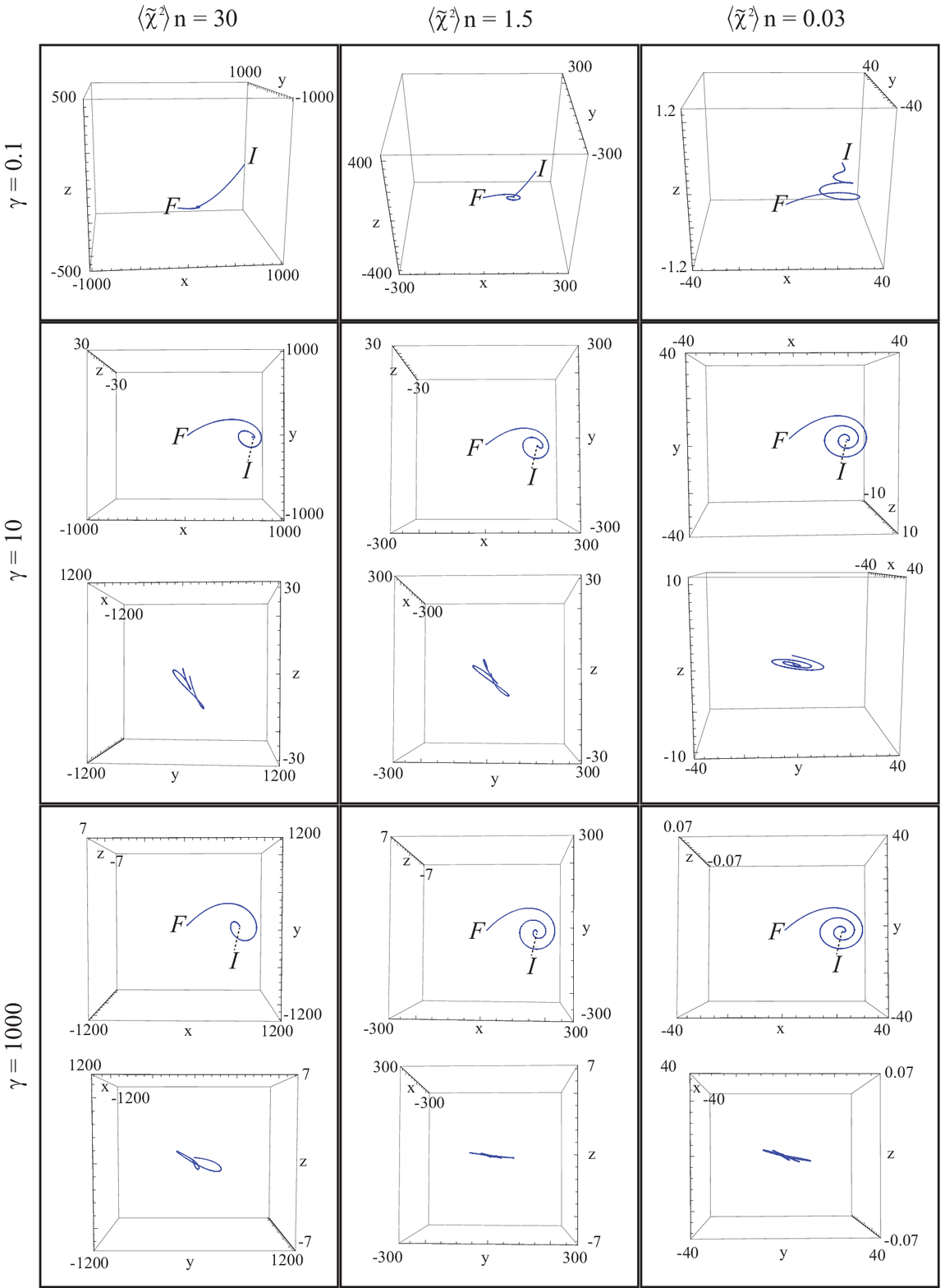}
  \caption{%
  Optimal paths corresponding to points in Fig.~\ref{fig:action} for various
values of
$n\langle\tilde\chi^2\rangle$ and $\gamma$ as indicated above
the columns and beside the rows respectively. The other parameters
are fixed at $A=20$ and $\chi=0$. In the bottom two rows each box shows an
optimal path from two
perspectives. Initial and final point are indicated by $I$ and $F$. Note that
the scales of the axes are typically different in different boxes. The left
column corresponds to the regime where detuning fluctuations dominate, the
right column to the regime where intrinsic fluctuations dominate and the center
column to the crossover.
  }\label{fig:optimalpaths}
\end{figure*}

We study the shapes of the optimal paths for the cases listed, with
representative examples given in Fig.~\ref{fig:optimalpaths}. Rather
surprisingly, we find that the shapes of the optimal paths are qualitatively
similar to the shapes of the deterministic paths, that are defined by
Eqs.~\eqref{eq:3deqsmotion}. Both the optimal path and the deterministic path
circle counter clockwise around the stable point $I$ while moving respectively
from/to the point. The three optimal paths in the right column of
Fig.~\ref{fig:optimalpaths}, with $n \langle\tilde\chi^2\rangle=0.03$,
correspond to case (i). In this case,
the intrinsic photon number fluctuations dominate. As a signature of this, the
path circles around the stationary point $I$ in the $xy$ plane, without any
significant displacement in the $z$ direction. 

The optimal paths in the column with $n \langle\tilde\chi^2\rangle=30$
correspond to the regime where the fluctuations in detuning dominate. Here the
displacements in the $z$ direction are two orders of magnitude larger than for
the optimal paths of the previous case. The upper left path corresponds to case
(iii), where the photon number fluctuations are sufficiently fast that the
optical
field adiabatically follows the fluctuations in the detuning. The path has a
shape resembling a piece of parabola and reaches
the saddle line close to the value of $z$ corresponding to the lasing
threshold
(Fig.~\ref{fig:lasing}). For the path with $\gamma=1000$
[case (iv)], detuning fluctuations occur on shorter time scales so that the 
photon number fluctuations become relatively more important, leading to an
increased number of circulations of the path around the stationary point $I$. 
The other optimal paths correspond to crossovers between the cases. The
general trend is however clear: With increasing $\gamma$ ($n
\langle\tilde\chi^2\rangle$) the displacement of the optimal path in the
$z$ direction ($xy$ plane) becomes smaller.

We briefly comment on the accuracy of the method used. The order of the Bezier
curve that is required to have an accurate optimal path is different for the
different cases. The simple optimal path of case (iii) is found with
high accuracy (up to
1\%) with a Bezier curve of the order $12$, while the optimal path with
$\gamma=1000$ and $n \langle\tilde\chi^2\rangle=1.5$ is a Bezier curve of order
$14$ and is less accurate (up to 10\%). We do not use a higher order
because it would greatly increase the computation time without providing new
insight. The shape of the optimal path can, however, be inferred, especially for
case (i). Since the
optimal path resembles the deterministic path and the deterministic
path in case (i) circles $\sim A/2\pi$ around the stable point
(Sec.\ref{sec:II}), we expect that also the optimal path circles $\sim A/2\pi$
times around $I$. To calculate this optimal path requires a Bezier curve of at
least the order of $A/2\pi$. The method is thus inefficient for very big $A$.
For the value $A=20\gg1$ that was used, we are far away from the lasing
threshold, while being able to find the optimal path with sufficient accuracy.
For the cases other than the case (i) the number of times the path circles
around $I$ decreases, resulting in more accurate optimal paths.

\section{Conclusions}\label{sec:V}
In this article we have shown that a device that combines stimulated emission
of light with superconductivity can produce coherent radiation with
exponentially long coherence times. 

We have introduced a general model for such HJL, where a
big number of quantum states are coupled to two superconducting leads
and are capable of emitting light by electron-hole recombination into a
resonant optical mode. We argued that there should also be incoherent
emission of photons in the environment that creates an imbalance in the
population of emitter states required for lasing. The stochastic dynamics of the
HJL is compactly described by a single Fokker-Planck equation,
Eq.~\eqref{eq:FPE}, for three real variables, incorporating two noise sources.
One noise source is the intrinsic quantum noise associated with the discrete
changes of photon number in the resonator mode. The other noise source is
related to the random switching between the eigenstates of the quantum emitters
and causes the fluctuations of detuning.

We have found a lasing regime where the nonlasing state is unstable. The optical
phase is locked to the superconducting phase difference in two
possible stable lasing states with a phase difference of $\pi$. In this
case, small fluctuations do not lead to decoherence, which is caused by a large
fluctuation whereby the device switches between the two stable lasing states.
To evaluate the coherence time of the HJL, we study these large fluctuations
estimating their probability by the optimal path method. Our numerical results
show agreement with simple estimations deduced from the distribution of small
fluctuations. 

We have found the decoherence times to be exponentially long $\simeq
\exp[-S_\text{opt}]$ provided $S_\text{opt}\gg 1$. The optimal action is
inversely proportional either to the number of photons in the laser or to the
variance of the detuning. The latter is inversely proportional to the number
$N$ of emitters in the laser. Thereby we prove the feasibility of the
exponentially long times for a big number of emitters. $S_\text{opt}$ was also
shown to depend on $\gamma$, the ratio of emitter switching rates to damping
rate from the resonant mode, and the distance to the lasing threshold.

To give an example we can express these conditions for a simple
microscopical model formulated in terms of the parameters of the
interaction Hamiltonians [Eqs.~\eqref{eq:ham_SC} and~\eqref{eq:ham_photon}] in
Sec.~\ref{sec:I}. First, we estimate the system parameters
$A$ and $\Omega''$ [Eq.~\eqref{eq:energy}] assuming that the energy of the
states of the emitters is only perturbed by
these two interactions. Taking $E_s$ as the typical energy
scale of the eigenstate of a single emitter and using perturbation theory, we
can
therefore estimate the
order of magnitude of $A$ and $\Omega''$
	\begin{align} 
	   A\simeq  N\frac{|\Delta_e\Delta_h|G^2}{E_s^3},\quad 
		\Omega''\simeq N \frac{G^4}{E_s^3}.
	\end{align}
Here it is assumed that the population of emitter states is asymmetric as
discussed in Sec.\ref{sec:I}. To have lasing in the HJL requires $A>\Gamma$.
Additionally, a large number of photons, $n$, in the resonator mode requires
$A\gg\Omega''$, implying that $|\Delta_e\Delta_h|\gg G^2$. Second, we estimate
the variance of the detuning fluctuations, $\tilde\chi$, which
correspond to the fluctuations of $\Omega'$ [Eq.~\eqref{eq:energy}] scaled with
$A$. Using perturbation theory, we estimate $\langle\Omega'\rangle\simeq
NG^2/E_s$. Its variance is of the order of $\alpha\langle\Omega'\rangle^2/N$,
with $\alpha\lesssim1$ being a constant depending on the transition rates
between the emitter eigenstates, where it is of the order of one when all rates
are of the same order. This leads to 
	\begin{align}
	   \langle\tilde\chi^2\rangle\simeq
			\alpha\frac{E_s^4}{N|\Delta_e\Delta_h|^2},
	\end{align}
where the estimation for $A$ was used. Typically, this quantity should be much
smaller than one.

We list several ideas for possible applications of the HJL.
In the HJL model, we neglected voltage fluctuations of the bias.
When present, these fluctuations will transfer to fluctuations in the
optical phase of the laser. By measuring these and providing negative feedback
the voltage fluctuations, and hence the optical fluctuations, can be reduced and
stabilized.
Another idea is similar to what was proposed in Ref.~\onlinecite{recher:10}. One
can embed two HJLs in a superconducting interference device (SQUID) and let them
interact optically, so that
an Aharonov-Bohm effect manifests in a shift of the optical phases that is
flux-dependent. 
As an alternative to the latter idea, one can also let two distant SQUIDs
interact
via optically coupled HJLs. This may allow for coherent interactions between
widely separated superconducting qubits. 
Finally, the exponentially long coherence time of the HJL can make it very
suitable for metrology purposes, where the measurement of optical frequencies in
the next-generation atomic clocks require ultrastable optical
lasers.\cite{metrology}

\acknowledgments

We acknowledge fruitful discussions with F. Hassler, C. Padurariu and C.J.O.
Verzijl and financial support from the Dutch Science Foundation NWO/FOM.

\appendix

\section*{Appendix: Noise spectrum}
The noise spectrum of Eq.~\eqref{eq:noisespectrum} can be written in three
parts, each part representing a peak. It is symmetric in $\nu$
and is given by
	\begin{align*}
	   S(&\nu) = \\
			&\frac{C_1\nu+C_2}{(\nu + \sqrt{D-1})^2+1} 
				+ \frac{-C_1\nu+C_2}{(\nu - \sqrt{D-1})^2+1} +
\frac{\gamma C_3}{\nu^2+\gamma^2} 
	\end{align*}
with $D = 4(A^2-1)(1+\chi)$. Defining $W = \gamma^4+2\gamma^2(D-2)+D^2$,
the coefficients are given by
\begin{widetext}
	\begin{align*}
	   C_1 &= \frac{1}{16DW\sqrt{D-1}}\Big\{ \left[ 4A^2+D(\chi-1) \right] W
		 + 32n_s\gamma\langle\tilde\chi^2\rangle (A^2-1)^2\left[ 
			4(D-1)+\chi(D-\gamma^2) \right] \Big\}, \\
		C_2 &= \frac{1}{16DW}\Big\{ 2W\left[ 4A^2+D(1+\chi) \right] 
		 + 64n_s\gamma\langle\tilde\chi^2\rangle (A^2-1)^2\left[ 
			\gamma^2 -4 + D(3+\chi) \right] \Big\}, \\
		C_3 &= \frac{1}{4W}8n_s\langle\tilde\chi^2\rangle
			(A^2-1)\left[4(A^2-1)-\gamma^2\right],
	\end{align*}
\end{widetext}


\begin{thebibliography}{10}

\bibitem{tinkham}
M. Tinkham,
 {\em Introduction to Superconductivity\/},
 2nd~edition (McGraw-Hill, New York, 1996).

\bibitem{scully}
M.O. Scully and M.S. Zubairy,
 {\em Quantum optics\/}
 (Cambridge University Press, Cambridge, 1997).

\bibitem{scullylamb}
M.O. Scully and W.E. Lamb, Phys. Rev. {\bf 159}, 208-226 (1967).

\bibitem{godschalk}
F. Godschalk, F. Hassler and Yu.V. Nazarov, Phys. Rev. Lett. {\bf 107}, 073901
(2011).

\bibitem{recher:10}
P. Recher, Yu.V. Nazarov and L.P. Kouwenhoven,
   Phys. Rev. Lett. {\bf 104}, 156802 (2010).

\bibitem{hassler}
F. Hassler, Yu.V. Nazarov and L.P. Kouwenhoven, Nanotechnology {\bf 21}, 274004
(2010).

\bibitem{vanDam&Doh}
Y.J. Doh, J.A. van Dam, A.L. Roest, E.P.A.M. Bakkers, L.P. Kouwenhoven and S.
De Franceschi, Science {\bf 309}, 272 (2005); J.A. van Dam, Yu.V. Nazarov,
E.P.A.M. Bakkers, S. De Franceschi and L.P. Kouwenhoven,
Nature {\bf 442}, 667 (2006).

\bibitem{spdc}
S.E. Harris, M.K. Oshman and R.L. Byer, Phys. Rev.
Lett. {\bf 18}, 732 (1967).

\bibitem{ohtsu}
M. Ohtsu, 
  {\em Highly Coherent Semiconductor Lasers\/}
  (Artech, Boston, Mass., 1992).

\bibitem{lang}
R. Lang, IEEE Journal of Quantum Electronics {\bf 18}, 976 - 983 (1982), and
references therein.

\bibitem{populationinversion} 
This population imbalance must not be confused with a population inversion which
is usually required for lasing in common lasers.

\bibitem{voltagestandard}
C.J. Burroughs, S.P. Benz, T.E. Harvey and C.A. Hamilton, IEEE Transactions on
Applied Superconductivity {\bf 9}, 4145-4148 (1999).

\bibitem{S-2DEG-S}
Y. Hayashi, K. Tanaka, T. Akazaki, M. Jo, H. Kumano and I. Suemune, Appl. Phys.
Express {\bf 1}, 011701 (2008);
H. Sasakura, S. Kuramitsu, Y. Hayashi, K. Tanaka, T. Akazaki, E.
Hanamura, R. Inoue, H. Takayanagi, Y. Asano, C. Hermannst\"adter, H.
Kumano and I. Suemune, Phys. Rev. Lett. {\bf 107}, 157403 (2011);
I. Suemune, H. Sasakura, Y. Hayashi, K. Tanaka,
T. Akazaki, Y. Asano, R. Inoue, H. Takayanagi,
E. Hanamura, J.-H. Huh, C. Hermannst\"adter, S. Odashima
and H. Kumano, Japanese Journal of Applied Physics {\bf 51}, 010114
(2012).

\bibitem{stackedjj}
V.M. Krasnov, Phys. Rev. Lett. {\bf 97}, 257003 (2006);
L. Ozyuzer, A. E. Koshelev, C. Kurter, N. Gopalsami, Q. Li, M. Tachiki,
K. Kadowaki, T. Yamamoto, H. Minami, H. Yamaguchi, T. Tachiki, K. E. Gray,
W.-K. Kwok and U. Welp, Science {\bf 318}, 1291 (2007);
T. M. Benseman, A. E. Koshelev, K. E. Gray, W.-K. Kwok, U. Welp,
K. Kadowaki, M. Tachiki and T. Yamamoto , Phys. Rev. B {\bf 84}, 064523 (2011).

\bibitem{diedrich}
F. Diedrich, J.C. Bergquist, W.M. Itano, and D.J. Wineland, Phys. Rev. Lett.
{\bf 62}, 403 (1989).

\bibitem{risken}
H. Risken, C. Schmid and W. Weidlich, Zeitschrift f\"ur Physik {\bf 194},
337-359 (1966).

\bibitem{vankampen}
N.G. van Kampen,
	{\em Stochastic Processes in Physics and Chemistry\/},
	3rd~edition (Elsevier, Amsterdam, 2007).

\bibitem{reviewop}
M.I. Freidlin and A.D. Ventzel, {\em Random perturbations in
dynamical systems} (Springer, Berlin, 1984); 
M.I. Dykman, E. Mori, J. Ross and P.M. Hunt, 
J. Chem. Phys. {\bf 100} (8), 5735-5750 (1994).

\bibitem{kramers}
H.A. Kramers, Physica {\bf 7}, 284 (1940).

\bibitem{kleinert}
H. Kleinert,
{\em Path Integrals\/}, 
2nd~edition (World Scientific, New Jersey, 1995).

\bibitem{merzbacher}
E. Merzbacher,
{\em Quantum Mechanics\/},
2nd~edition (John Wiley and Sons Inc, New York, 1970).

\bibitem{gardiner}
C.W. Gardiner, Journal of Statistical Physics {\bf 30},
157-177 (1983).

\bibitem{metrology}
T. Kessler, C. Hagemann, C. Grebing, T. Legero, U. Sterr, F. Riehle, M. J.
Martin, L. Chen and J. Ye, Nature Photonics {\bf 6}, 687-692 (2012);
F. Riehle, Physics {\bf 5}, 126 (2012).


\end{thebibliography}
\end{document}